\documentclass[aps,prl,twocolumn,superscriptaddress,showpacs,footnotebib]{revtex4-1}
\usepackage{amsmath}
\usepackage{bm}
\usepackage{color}
\usepackage{graphicx}
\usepackage{epstopdf}
\usepackage{epsfig}
\usepackage{amsfonts}
\usepackage[naturalnames]{hyperref}
\usepackage{hypcap}
\usepackage{verbatim}
\usepackage{tabularx}
\usepackage{bbm}
\usepackage{esvect}
\usepackage{subfigure}
\usepackage{slashed}

\usepackage{slashed}

\newcommand{\dk}[1]{\left( #1 \right)}

\usepackage[T1]{fontenc} 

\def\bea{\begin{eqnarray}}
\def\eea{\end{eqnarray}}

\def\ba{\begin{array}}
\def\ea{\end{array}}

\hypersetup{colorlinks=true, citecolor=blue, urlcolor=blue, linkcolor=blue}

\begin{document}

\title{Boundary and defect criticality in topological insulators and superconductors}

\author{Xiaoyang Shen}
\affiliation{Department of Physics, Tsinghua University, Beijing 100084, China}
\affiliation{Institute for Advanced Study, Tsinghua University, Beijing 100084, China} 
\author{Zhengzhi Wu}
\affiliation{Institute for Advanced Study, Tsinghua University, Beijing 100084, China}
\author{Shao-Kai Jian}
\email{sjian@tulane.edu}
\affiliation{Department of Physics and Engineering Physics, Tulane University, New Orleans, Louisiana, 70118, USA}

\begin{abstract}
We study the boundary criticality enriched by boundary fermions, which ubiquitously emerge in topological phases of matter, with a focus on topological insulators and topological superconductors. 
By employing dimensional regularization and bosonization techniques, we uncover several unprecedented boundary universality classes.
These include the boundary Gross-Neveu-Yukawa critical point and the special Berezinskii-Kosterlitz-Thouless (BKT) transition, both resulting from the interplay between edge modes and bulk bosons.   
We present a comprehensive sketch of the phase diagram that accommodates these boundary criticalities and delineate their critical exponents.
Additionally, we explore a 1+1D conformal defect decorated with fermions, where a defect BKT transition is highlighted. 
We conclude with a discussion on potential experimental realizations of these phenomena.
\end{abstract}
\maketitle

{\it Introductions.---} Boundary physics holds a pivotal role across diverse fields, such as the quantum Hall effect with boundary states in condensed matter physics~\cite{vonKlitzing1980new,tsui1982two,laughlin1981quantized,laughlin1983anomalous} and the AdS/CFT correspondence~\cite{maldacena1997the,Witten1998anti,gubser1998gauge} involving quantum field theory at the boundary of a gravitational system in string theory \cite{polchinski1996TASI,takayanagi2011holographic,Fujita_2011,Recknagel:2013uja,Izumi_2022}. 
For critical phenomena, the renormalization group (RG) flow leads to universality classes, where the system exhibits scale invariance. 
In this context, the presence of boundaries enriches the universality class, giving rise to unique boundary conformal field theories (BCFTs)~\cite{cardy1984conformal,cardy1989boundary,affleck1991universal}. 
BCFT serves as a unified framework for understanding boundary critical physics across diverse physical systems. 
Hence, investigations into BCFT hold a broad appeal and bear wide-ranging applications across numerous research disciplines~\cite{cardy2008boundaryconformalfieldtheory,Ge:2024ubs,Shen:2023srk,andrei2018boundarydefectcftopen,ji2024topologicaldefects21dsystems,brillaux2021fermi,brillaux2024surface,myerson2024pristine,ma2022edge}. 

An important research field in condensed matter physics that is closely related to boundaries is the symmetry protected topological state (SPT) \cite{kitaev2009periodic,ryu2010topological,Gu_2009,hasan2010toopological,Qi_2011,chen2012symmetry}. 
The distinct feature of SPT from a trivial state is the presence of gapless boundary modes. 
A well-known example is the topological insulator~\cite{kane2005quantum,bernevig2006quantum}, which is characterized by gapless fermions on the boundary. 
Because the topological insulator has been realized experimentally~\cite{molenkamp2007quantum,hsieh2008topological}, it is practically important to investigate the interplay between BCFTs and the SPT~\cite{zhang2017unconventional,wu2020boundary,ma2022edge}. 
Notice that the notation of SPT has recently been generalized to the realm of gapless phases of matter~\cite{scaffidi2017gapless,verresen2020topology,verresen2021gapless,yu2022conformal,yu2024universal}. 
Nevertheless, the higher dimensional cases involving weakly interacting fermionic SPT states have not been fully explored yet, especially in a theoretically controlled way. 

In this paper, we study the boundary criticalities enriched by boundary gapless fermionic modes~\cite{Liu:2021nck,2022Giombia,2023Giombib,2023Barrat,Shachar:2024ubf,Herzog_2023}, with an application to the topological insulators (TIs) and topological superconductors (TSCs) protected by the time-reversal symmetry. 
It is well-known that in the presence of a boundary, the Wilson-Fisher (WF) fixed point can be enriched into ordinary, extraordinary, or special transition depending on the coupling strength at the surface layer~\cite{1981Diehl,1983Diehl,2020Giombi,2022Padayasi,2022Parisen,2022metlitski}. 
A new ingredient in our model is that a gapless mode, a Dirac fermion or Majorana fermion, resides on the boundary, and enriches the transition via a Yukawa coupling at the boundary. 
By implementing a $4-\epsilon$ RG calculation, we uncover a boundary Gross-Neveu-Yukawa fixed point at the boundary of TSCs. 
On the other hand, in the case of 2+1D TIs, because the four-fermion interaction is marginal at its 1+1D edge, we include it nonperturbatively by bosonization.
In stark contrast to the boundary transition in the $O(N)$ model, we find an unconventional ordinary transition enriched by Luttinger liquid with a continuous boundary scaling dimension. 
Interestingly, this unconventional ordinary transition terminates at a Berezinskii–Kosterlitz–Thouless (BKT) like transition~\cite{kosterlitz2018ordering,Berezinsky:1970fr}, which is termed as the special BKT transition. 
We also extend our bosonization calculation to investigate the 1+1D conformal defect~\cite{Chang:2022hud,2023Simone,Trepanier:2023tvb} decorated with fermions, and find a parallel scenario, where a distinct conformal defect with a continuous scaling dimension terminates at a defect BKT transition.
Finally, we briefly discuss the experimental probes of various boundary criticalities, and propose possible experimental realizations.

\begin{figure*}[t]
    \centering        
    \subfigure[]{\includegraphics[height = 0.28\textwidth]{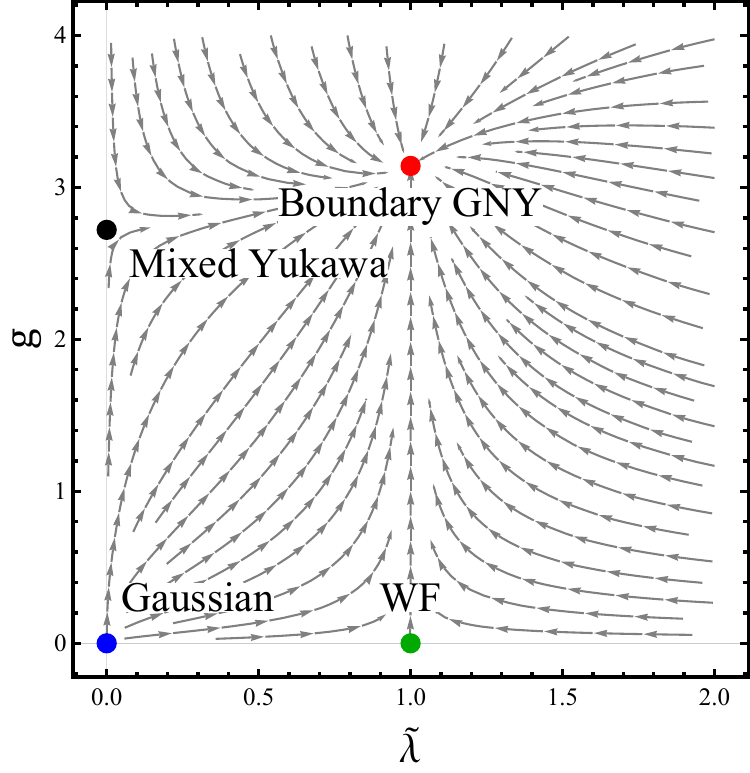}} \qquad 
    \subfigure[]{\includegraphics[height = 0.28\textwidth]{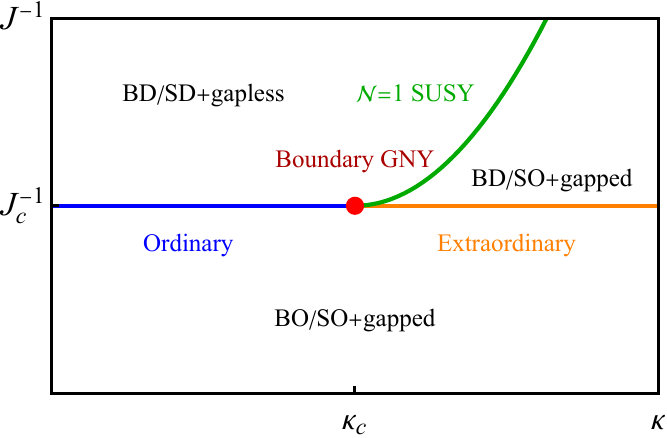}}
    \caption{ (a) RG flow  parametrized by $\tilde{\lambda}$ and $g$. $\tilde{\lambda}$ stands for $3\lambda/(4\pi)^2$. (b) Phase diagram of boundary criticality in TSCs as functions of the bulk coupling and the boundary enhancement. 
    BD(BO)/SD(SO)+gapless/gapped stands for bulk disordered (bulk ordered)/surface disordered (surface ordered) with gapless/gapped boundary fermion. 
    The multicritical point corresponds to the boundary Gross-Neveu-Yukawa fixed point illustrated in (a). }
    \label{fig:rg_flow}
\end{figure*}

{\it Boundary Ising criticality.---} We briefly review the boundary Ising critical theory. 
The field theory of an Ising model in a $d$-dimensional Euclidean spacetime $\mathcal M$ with a boundary $\partial \mathcal M$ is given by
\begin{equation} \label{eq:ising_boundary}
    S_\text{b} = \int_{\mathcal M} d^d x \left[\frac{1}{2}(\partial \phi)^2+\frac{\lambda}{4!}\phi^4\right] + \int_{\partial \mathcal M} d^{d-1}x h\phi^2\,,
\end{equation}
where $\phi$ denotes a real scalar field, $(\partial \phi)^2 = \sum_{\mu=0}^{d-1}(\partial_\mu \phi)^2 $ and $\lambda$ is the coupling strength. 
The second term denotes a surface term, with $h$ being a surface mass~\footnote{Some boundary symmetry-breaking coupling that drives the system to the normal universality class is neglected. See Diehl \cite{1981Diehl,1983Diehl} for a detailed discussion.}. 
The bulk is at the WF fixed point, and in the RG calculation, it does not receive corrections from the surface.
Hence, depending on the surface mass $h$, there are distinct boundary universality classes: 
the special transition corresponds to vanishing surface mass $h = 0$~\footnote{Strictly speaking, the surface mass is located at some $h=h_\text{sp}$ for the special transition.}, and the ordinary and extraordinary transition correspond, respectively, to a disordered and ordered surface, together with $h\to +\infty$ and $h\to -\infty$. 

{\it Boundary GNY criticality.---} Our model starts by introducing the fermion, $\psi$, on the boundary, coupling to the surface order parameter,
\begin{equation} \label{eq:action_general_spt}
    \begin{split}
    S_\text{bGNY} =& \int_{\mathcal M} d^{d} x \left[ \frac12 (\partial \phi)^2 + \frac{\lambda}{4!}  \phi^4\right] \\
    & + \int_{\partial \mathcal M} d^{d-1}x \left( \bar\psi \slashed \partial \psi + g \phi \bar \psi \psi +h\phi^2 \right)\, ,
    \end{split}
\end{equation}
where $g$ denotes the boundary Yukawa coupling, and $\psi$ denotes a two-component surface Majorana fermion of a TSC or a two-component surface Dirac fermion of a TI.  
We use the convention: $\bar\psi=\psi^{\dagger}\gamma^0$ ,$\slashed \partial=  \sum_{\mu=0}^2 \partial_{\mu}\gamma^{\mu}$, and $\gamma^0=\sigma_z, \gamma^1=-\sigma_y,\gamma^2=\sigma_x$. 
Since we are interested in the BCFT, the bulk is tuned to the WF critical point with $\phi$ being the Ising order parameter field.
When the Ising field $\phi$ orders, it will break the time reversal symmetry and consequently gap out the edge mode $\psi$. 
In general, the boson velocity $v_b$ and fermion velocity $v_f$ can be different. 
However, it can be shown that there exists a stable fixed point $(v_f/v_b)_* = 1$, implying an emergent Lorentz invariance at the boundary~(see Supplemental Material Sec. I). 
The emergent Lorentz symmetry allows us to scale $v_f = v_b = 1$.

At the special transition $h = 0$, dimension regularization with $\epsilon = 4-d$ leads to the following the RG equations~(see Supplemental Material Sec. II),
\begin{eqnarray} \label{eq:bGNY}
    \frac{dg}{dl} & = & \frac{\epsilon}{2}g-\frac{2}{3\pi^2}g^3+\frac{1}{32\pi^2}\lambda g \,,\\
\label{eq:ising_bulk}       
    \frac{d\lambda}{dl} & = &\epsilon\lambda-\frac{3}{16\pi^2}\lambda^2 \,.
\end{eqnarray}
Note that the second equation is exactly the 1-loop RG equation for the bulk WF fixed point because the bulk does not receive renormalization from the boundary. 
The bulk coupling, however, renormalizes the boundary Yukawa coupling. 
The RG equations have four types of fixed points: a UV stable Gaussian fixed point, a WF fixed point, which is unstable along $g$ direction, a mixed Yukawa fixed point with $\lambda_* = 0$ and $g_*^2 = \frac{3\pi^2\epsilon}{4}$~\cite{2017Christopher}, which is unstable along $\lambda$ direction, and an IR stable boundary Gross-Neveu-Yukawa (GNY) fixed point with $\lambda_* = \frac{(4\pi)^2\epsilon}{3}$ and $g_*^2 = \pi^2\epsilon$. 
The RG flow is shown in Fig.~\ref{fig:rg_flow} (a).
At the boundary GNY fixed point, the critical exponent for the boundary order parameter and the boundary correlation length are unchanged, i.e., $\tilde \Delta_{\hat{\phi}}= \frac{d-2}2 - \frac{\epsilon}{6}$, $\frac1{\nu_\text{bdy}} =2 - \frac{\epsilon}3$, but most interestingly, we obtain a new anomalous dimension of the fermion, $\Delta_{\psi} = \frac{d-1}2 + \frac{\epsilon}{12}$.

The above perturbative $\epsilon$ expansion is trustworthy for small $\epsilon$.
We believe the boundary GNY fixed point can be extrapolated to $\epsilon = 1$ for a 2+1D bulk system with a 1+1D edge in the context of topological superconductivity.
More precisely, it describes the special transition at a spontaneous time reversal symmetry breaking transition of a 2+1 dimensional topological superconductor.
In this case, the time reversal symmetry protects gapless edge Majorana fermions, which are coupled to the boundary order parameter via a boundary Yukawa coupling.
When the time reversal symmetry is spontaneously broken in the bulk, the gapless edge fermion is no longer protected. 
Based on the phase diagram of the boundary critical Ising model, one can sketch the phase diagram enriched by the edge mode, as shown in Fig.~\ref{fig:rg_flow} (b). 
The phase diagram is controlled under two parameters, the bulk coupling $J$ and the ratio between the surface coupling $J_1$ and bulk coupling $\kappa \equiv J_1/J$, sometimes dubbed as the surface enhancement. 
Consider a weak bulk coupling $J<J_c$ and a weak surface enhancement $\kappa < \kappa_c$, both bulk and surface are disordered.
As the bulk coupling increases to $J > J_c$, the bulk and surface order simultaneously, and the ordinary transition emerges at $J = J_c$. 
At the ordinary transition, thanks to the Dirichlet boundary condition, the boson gradient $\partial_z \phi$ couples to the gapless boundary fermions, and the coupling turns out to be irrelevant.
As a result, the fermion is a spectator at the ordinary transition.
On the other hand, for a large surface enhancement $\kappa > \kappa_c$, the critical value of the coupling at the surface is less than that in the bulk. 
In this case, starting from a weak bulk coupling $J<J_c$, as the bulk coupling increases, the surface will order with the bulk remaining disordered.
At this surface transition point, one can utilize the surface order parameter $\hat \phi$ and edge Majorana mode $\psi$ to construct an effective field theory solely on the boundary~\footnote{For clarity, we emphasize that the terminology ``boundary criticality'' in this paper refers to the criticality that originates from interplay between different dimensional degrees of freedom, i.e. gapless bulk bosons and gapless edge modes in the context of topological insulator and superconductor.
The theory constructed solely via boundary degrees of freedom at the surface transition does not fall into this category.}. 
The theory then flows to the celebrated fixed point with an emergent $\mathcal{N} = 1$ supersymmetry or tricritical Ising critical point ~\cite{friedan1984superconformal,Grover:2013rc,jian2015,Fei:2016sgs,li2017edge}. 
As the bulk coupling keeps increasing and reaches the critical strength of the bulk, an extraordinary phase transition takes place.
Since the fermion is gapped, it is not involved in the extraordinary transition. 
Now, most interestingly, the ordinary transition and extraordinary transition merge at a multicritical point, $J = J_c$ and $\kappa = \kappa_c$, the special fixed point. 
As we discussed above, the special fixed point decorated by the edge fermions is described by the boundary GNY fixed point.

\begin{figure}
    \centering
    \includegraphics[width=0.8\linewidth]{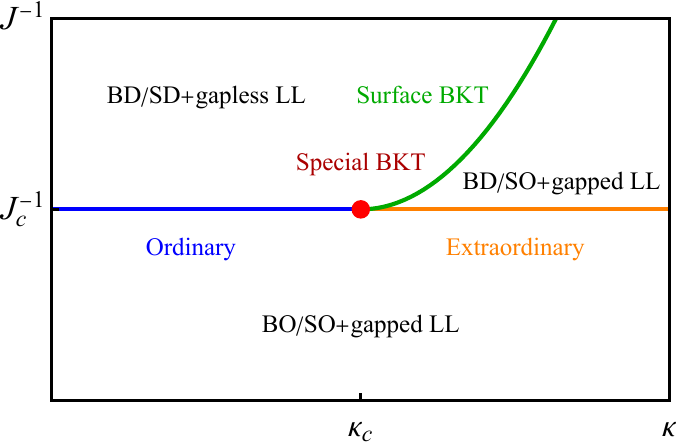}
    \caption{Phase diagram for boundary criticality in TIs as a function of the bulk coupling $J$ and the boundary enhancement $\kappa$. Note that the ordinary transition is also enriched by gapless boundary fermions.}
    \label{fig:phase_diagram_TI_boundary}
\end{figure}

\begin{table*}
    \centering
    \def\arraystretch{1.3}
    \begin{tabular}{c| c|c | c } 
    \hline
    \hline
    Boundary universality class & $\Delta_{\psi}$ & $\tilde \Delta_{\hat \phi} $ & $1/\nu_\text{bdy} $ \\
    \hline
    \hline
    Boundary GNY ($\epsilon=4-d$) & $\frac{d-2}2 + \frac\epsilon{12}$ & $\frac{d-2}2 - \frac\epsilon6$ & $2 - \frac{\epsilon}3 $  \\
    \hline
    Special BKT ($K_c = 2- \Delta_\text{ord}$)  & $\frac12 + \frac14 \left( K_c + \frac1{K_c} -2 \right) $ & $K_c$ & $\xi \propto e^{\frac{b}{\sqrt{K - K_c}}} $ \\
    \hline
    Defect BKT ($K_c' = 2- \Delta_\text{defect}$)  & $\frac12 + \frac14 \left( K_c' + \frac1{K_c'} -2 \right) $ & $K_c'$ & $\xi \propto e^{\frac{b'}{\sqrt{K - K_c'}}} $   \\
    \hline
\end{tabular}
\caption{Critical exponents at the boundary criticality in topological systems. 
The $\Delta_\text{ord}$ ($\Delta_\text{defect}$) denotes the scaling dimension of the boundary (defect) order parameter. 
$\tilde \Delta_{\hat \phi}$ denotes the scaling dimension in the fermion enriched boundary transition. 
$\Delta_\text{ord} = 1.2626$ from Monte Carlo simulation~\cite{deng2005surface} in $d=2+1$, and $\Delta_\text{defect} = 1 - \frac\epsilon6$ from a $d = 4-\epsilon$ calculation~\cite{2023Simone}. 
$b,b'$ are nonuniversal constants and $d$ is the bulk dimension.}
\label{tab:exponent}
\end{table*}

{\it Special BKT transition.---} While the boundary GNY fixed point can be realized in the transition in TSCs, for 2+1D TIs protected by time-reversal symmetry, a crucial issue arises at its boundary, i.e., four-fermion interactions are marginal and consequently should be considered~\footnote{Four fermion interaction, such as $\int d^2 x (\psi^\dag \partial \psi)^2$, is irrelevant for the Majorana fermion}.
To include the effect of four-fermion interactions, instead of a $4-\epsilon$ RG calculation, we analyze the boundary theory directly in 1+1D by bosonization~\cite{2013Fradkin,wu2006}. 
More precisely, there are two marginal four-fermion interactions respecting time-reversal symmetry, the forward-scattering and the two-particle backward-scattering interactions permitted at half-filling~\cite{wu2006}.
The former can be lumped non-perturbatively into the Luttinger parameter, while the latter is presented by a vortex operator. 
The boundary theory after bosonization is (see Supplemental Material Sec. III)
\begin{equation}
    \begin{split} \label{eq:bosonization}
    S_{\text{f}} =& \int d^2 x\Big[\frac{1}{2 K}(\partial \varphi)^2+g_1\phi \cos (\sqrt{4 \pi} \varphi) \\
    &+g_2 \cos\left(\sqrt{16 \pi} \varphi\right) 
    \Big] \,,
    \end{split}
\end{equation}
where $\varphi$ ($\phi$) denotes the bosonization field (the boundary order parameter), $K$ and $g_2$ present the Luttinger parameter and the strength of two-particle backward scattering, respectively, and $g_1$ is originated from the Yukawa coupling between the fermion and boson. 

Again, we consider the bulk boson is at the WF fixed point.
To proceed, we can hypothetically decouple the bulk boson from the boundary fermion at the outermost layer, i.e., $g_1 = 0$.
We expect this hypothetical bulk system without the outermost boundary layer to be described by the ordinary boundary universality class. 
We then apply the conformal perturbation theory at this ordinary transition fixed point for $g_1 \ne 0$. 
Because the boundary fermion will not affect the bulk RG flow, the relevant RG equations, consisting of only boundary couplings, yield 
\begin{equation}
\label{eq:boundary_bosonization}
    \begin{aligned}
        &\frac{dK}{dl}=-\pi^2 g_1^{2}K^2-4\pi^2g_2^2K^2 \,,\\
        &\frac{dg_1}{dl}=(2-\Delta_{\hat{\phi}}-K)g_1 \,, \\
        & \frac{dg_2}{dl}=(2-4K)g_2 \,,
    \end{aligned}
\end{equation}
where $\Delta_{\hat{\phi}} = \Delta_{\text{ord}}$ is the scaling dimension of the boundary order parameter for ordinary transition without fermions.
Without coupling to bulk bosons, i.e., $g_1 = 0$, the first and the third RG equations describe a BKT transition at $K  =1/2$ for the boundary fermion~\cite{wu2006,li2017edgeb}.
This can be understood as a conventional BKT transition: the repulsive four fermion interaction increases, which reduces $K$ from its noninteracting value $K=1$, then as the interaction strength exceeds a critical value, $K < 1/2$, it triggers a run-away flow for $g_2$ that condenses a boundary order. 

Interestingly, with a nontrivial coupling between the gapless bulk boson and the boundary fermion, i.e., $g_1 \ne 0$, the second RG equation is the new input for the boundary criticality. 
It indicates that when $K \le K_c \equiv 2 - \Delta_\text{ord} $, the coupling $g_1$ becomes relevant.
Hence, it is important to compare these two critical couplings $K = 1/2$ and $K_c = 2- \Delta_\text{ord} $.
A relatively accurate estimate of surface scaling dimension at the ordinary transition from Monte Carlo simulation is $\Delta_\text{ord} \approx 1.2626$~\cite{deng2005surface}. 
Hence, using this value, the critical strength is $K_c \approx 0.7374 > 1/2$, indicating a different BKT transition that happens at a weaker interaction strength. 
Actually, in the region $K < K_c$ where $g_1$ is relevant, it is natural to ask whether the theory flows to a gapped phase or a strongly coupled fixed point. 
To capture the behavior at the strong coupling region precisely, we perform the RG calculation up to the third order~\cite{Amit:1979ab}. 
The higher order RG flow shows that it flows to the massive phase instead of the strongly coupled fixed point (see Supplemental Material Sec. IV).

We can clearly understand how the edge state enriches the boundary criticality when the bulk is at the WF fixed point: 
At a weak boundary interaction, $K > K_c$, the presence of a Luttinger liquid at the edge renders a continuous scaling dimension for the surface order parameter $\tilde \Delta_{\hat \phi} = K$.
Note that we use $\tilde \Delta_{\hat \phi}$ to denote the scaling dimension of the boundary order parameter, i.e., $\cos (\sqrt{4\pi} \varphi)$, at the fermion enriched boundary transition. 
This is an unconventional ordinary transition enriched by the gapless edge state.
As the interaction increases to $K = K_c$, a BKT transition takes place, and subsequently, as it keeps increasing, i.e., $K < K_c$, the boundary boson orders spontaneously and the edge fermion is gapped out.
Notice that for $K < K_c$, the theory is described by the Ising extraordinary transition because the edge state is gapped. 
Hence, $K = K_c$ marks a special transition separating the unconventional ordinary transition and the Ising extraordinary transition, so we term it special BKT transition.
The critical exponents at this special BKT transition are given by $\tilde \Delta_{\hat \phi} =  K_c$, $\Delta_\psi = \frac12 + \frac14 \left( K_c + \frac1{K_c} - 2 \right)$, and the correlation length scales as $\xi \propto \exp\left[ \frac{b}{\sqrt{K - K_c}} \right] $, where $b$ is a nonuniversal number, as $K$ approaches the critical point $K_c$. 
With this understanding, the phase diagram is sketched in Fig.~\ref{fig:phase_diagram_TI_boundary}.

{\it Defect criticality in topological systems.---}
We extend our calculation to topological phases with a 1+1D defect \cite{2022metlitski,Chang:2022hud,2023Simone,raviv-moshe2023phases,Trepanier:2023tvb}. 
We briefly review the conformal defect in an Ising theory.
The action with an 1+1D mass defect in a $d$ dimensional critical Ising theory reads
\begin{equation} \label{eq:ising_defect}
    S_\text{d} = \int_{\mathcal M} d^dx \left[ \frac12 (\partial \phi)^2 + \frac{\lambda}{4!}  \phi^4 \right] + \int_{\mathcal D} d^2x h\phi^2 \,,
\end{equation}
where $h$ in the second term denotes the defect mass. 
Instead of a codimensional one boundary described in Eq.~\eqref{eq:ising_boundary}, it describes of a 1+1D defect, $\mathcal D$, embedded in a $d$ dimensional Euclidean spacetime $\mathcal M$. 
In $4- \epsilon$ regularization, the RG equation for $\lambda$ is the same as Eq.~\eqref{eq:ising_bulk}, while the RG equation for $h$ reads~\cite{2023Simone}
\begin{equation}
    \frac{d h}{d l} =\epsilon h-\frac{h^2}{\pi}-\frac{\lambda h}{16 \pi^2}\,,
\end{equation}
which reveals a nontrivial conformal defect at a stable fixed point $h_* = \frac{2\pi\epsilon}{3}$. 
At this fixed point, the scaling dimension of the defect order parameter $\Delta_{\text{defect}} = \frac{d-2}2+\frac{h_*}{\pi} = 1 - \frac{\epsilon}6 $ .

We assume that a gapless fermion is located at the defect, and couples to the defect order via a Yukawa term, i.e.,  $S_\text{f} = \int_{\mathcal D} d^2x \left(  \bar\psi \slashed \partial \psi + g \phi \bar \psi \psi \right)$. 
The convention of Dirac matrix is the same as in Eq.~\eqref{eq:action_general_spt}.  
We then employ the bosonization technique to study the effect of a decorated fermion on the conformal defect. 
The bosonized action is the same as Eq.~\eqref{eq:bosonization}, in which a marginal four-fermion interaction is included via a Luttinger parameter $K$.
The theory leads to the same RG equations as listed in Eq.~\eqref{eq:boundary_bosonization}, except now $\Delta_{\hat \phi} = \Delta_{\text{defect}}$. 
It implies a defect BKT transition at $K_c' = 2 -\Delta_\text{defect} = 1 + \frac{\epsilon}{6}$, where we use the result of the $4-\epsilon$ calculation for $\Delta_\text{defect}$ in the last equality. 
Crucially, the critical Luttinger parameter $K_c'$ is again greater than $1/2$, i.e., $K_c'  > \frac12$, for $\epsilon \le 1$,  which in turn indicates that the critical interaction strength for the defect BKT transition is weaker than the critical strength for local interactions. 

Hence, there is a parallel scenario for the fermion enriched conformal defect, in which the bulk is always located at the WF fixed point:
At weak interactions, $K>K_c'$, the conformal defect is enriched to have a continuous scaling dimension $\tilde \Delta_{\hat \phi} = K$ before a defect BKT transition taking place at $K = K_c'$.
The critical exponents at this defect BKT transition are summarized in Table~\ref{tab:exponent}. 
After the BKT transition, the Ising field orders on the defect and the fermion opens up a gap. 
The defect is then described by a normal universality class~\cite{2023Simone}.

It is also direct to analyze the defect criticality in the TSCs in the same fashion as the boundary case. 
Notice that the defect becomes an interface at $2+1$D bulk. 
The special interface fixed point can be identified by directly performing the standard dimensional regularization in the co-dimension one interface. 

{\it Concluding remarks.---} Possible experimental probes of the novel boundary criticality include scanning tunneling microscopy (STM)~\cite{BINNIG1983236,Binnig1987,eggert2000} and transport techniques \cite{1992kane,li2015}. 
Since the experimental realization of topological superconductors remains less well understood, we focus on topological insulators. 
The ordinary transition enriched by Luttinger liquid possesses an unconventional Yukawa term.
This can be observed in a transport experiment via a subleading correction $\delta G$ to the quantized conductance $G_0$~\cite{Landauer1995,maslov1995,lezmy2012}, $G = G_0-\delta G$.
The subleading contribution respects the scaling law,
$\delta G \sim E^{2\Delta_\text{Yukawa}-2} + E^{2\Delta -2}$, where the energy scale $E$ is given by the larger of either temperature or voltage, and $\Delta_\text{Yukawa} = \Delta_\text{ord} + K$ and $\Delta = 4K$. 
As $\Delta_\text{Yukawa} < \Delta$ in the ordinary transition, the Yukawa scaling can be extracted from the leading term in $\delta G$. 
Moreover, regarding the special BKT transition, a key feature distinct from the conventional one is the critical Luttinger parameter, $K_c$. 
This can be observed in differential tunneling conductance using STM, $\frac{dI}{dV} \sim E^{2\Delta_\psi-1}$.
These electronic experiments enable probing of boundary degrees of freedom, offering a novel approach for experimentally measuring boundary criticality that was previously inaccessible in the study of spin systems. 
Finally, given that boundary criticality is an interacting effect and involves the breaking of time-reversal symmetry, possible material candidates include topological Kondo insulators, such as SmB$_6$~\cite{dzero2010topological,alexandrov2013cubic,dzero2016topological}, and magnetic topological insulators, such as MnBi$_2$Te$_4$~\cite{otrokov2019prediction,bernevig2022progress}.

To conclude, we provide a comprehensive study of boundary and defect criticality in topological phases. 
We identify several novel boundary criticalities summarized in Table~\ref{tab:exponent}, and sketch the phase diagrams.
The fermion mode is treated as a perturbation to the original fixed point. 
Hence, it remains an interesting future question to see how the boundary criticality is modified in the strong coupling region.  
Moreover, there still exist many intriguing problems regarding the fermion-enriched boundary criticality. 
For instance, it would be valuable to investigate the higher-order boundary criticality, which can be realized in the higher-order topological insulators~\cite{benalcazar2017quantized, Schindler_2018}, and to investigate the boundary criticality in topological semimetals~\cite{2022Giombia,brillaux2021fermi,brillaux2024surface}. 
Besides, the study of the boundary criticality with $N>1$ flavor boson would be important, as some novel extraordinary-log transition may take place for a general value of $N$~\cite{Harribey:2023xyv,Sun:2023vwy,2022metlitski} and it is feasible to realize a fermion-enriched extraordinary-log universality in topological phase. 
In addition, the boundary criticality with gapless bulk and topological edge modes studied here can be viewed as a symmetry-enriched quantum critical point (QCP)~\cite{verresen2021gapless} in higher dimensions~\footnote{A discussion of the boundary Gross-Neveu-Yukawa class in 3+1 dimensions is provided in the Supplemental Material, where we obtain a new logarithmic correction to the boundary fermion correlation function.}.   
It is distinct from the conventional QCP in the Ising theory, requiring the existence of another quantum (multi-)critical point separating them.

{\it Acknowledgments.}
We thank Chi-Ming Chang, Xie Chen and Yang Ge for helpful discussions. 
The work is supported in part by a startup grant from Tulane University. 
Z. Wu acknowledges the support from the Shuimu Fellow Foundation at Tsinghua University.

\bibliography{reference}

\begin{thebibliography}{97}%
\makeatletter
\providecommand \@ifxundefined [1]{%
 \@ifx{#1\undefined}
}%
\providecommand \@ifnum [1]{%
 \ifnum #1\expandafter \@firstoftwo
 \else \expandafter \@secondoftwo
 \fi
}%
\providecommand \@ifx [1]{%
 \ifx #1\expandafter \@firstoftwo
 \else \expandafter \@secondoftwo
 \fi
}%
\providecommand \natexlab [1]{#1}%
\providecommand \enquote  [1]{``#1''}%
\providecommand \bibnamefont  [1]{#1}%
\providecommand \bibfnamefont [1]{#1}%
\providecommand \citenamefont [1]{#1}%
\providecommand \href@noop [0]{\@secondoftwo}%
\providecommand \href [0]{\begingroup \@sanitize@url \@href}%
\providecommand \@href[1]{\@@startlink{#1}\@@href}%
\providecommand \@@href[1]{\endgroup#1\@@endlink}%
\providecommand \@sanitize@url [0]{\catcode `\\12\catcode `\$12\catcode
  `\&12\catcode `\#12\catcode `\^12\catcode `\_12\catcode `\%12\relax}%
\providecommand \@@startlink[1]{}%
\providecommand \@@endlink[0]{}%
\providecommand \url  [0]{\begingroup\@sanitize@url \@url }%
\providecommand \@url [1]{\endgroup\@href {#1}{\urlprefix }}%
\providecommand \urlprefix  [0]{URL }%
\providecommand \Eprint [0]{\href }%
\providecommand \doibase [0]{http://dx.doi.org/}%
\providecommand \selectlanguage [0]{\@gobble}%
\providecommand \bibinfo  [0]{\@secondoftwo}%
\providecommand \bibfield  [0]{\@secondoftwo}%
\providecommand \translation [1]{[#1]}%
\providecommand \BibitemOpen [0]{}%
\providecommand \bibitemStop [0]{}%
\providecommand \bibitemNoStop [0]{.\EOS\space}%
\providecommand \EOS [0]{\spacefactor3000\relax}%
\providecommand \BibitemShut  [1]{\csname bibitem#1\endcsname}%
\let\auto@bib@innerbib\@empty
\bibitem [{\citenamefont {von Klitzing}\ \emph {et~al.}(1980)\citenamefont {von
  Klitzing}, \citenamefont {Dorda},\ and\ \citenamefont
  {Pepper}}]{vonKlitzing1980new}%
  \BibitemOpen
  \bibfield  {author} {\bibinfo {author} {\bibfnamefont {K.}~\bibnamefont {von
  Klitzing}}, \bibinfo {author} {\bibfnamefont {G.}~\bibnamefont {Dorda}}, \
  and\ \bibinfo {author} {\bibfnamefont {M.}~\bibnamefont {Pepper}},\ }\href
  {\doibase 10.1103/PhysRevLett.45.494} {\bibfield  {journal} {\bibinfo
  {journal} {Phys. Rev. Lett.}\ }\textbf {\bibinfo {volume} {45}},\ \bibinfo
  {pages} {494} (\bibinfo {year} {1980})}\BibitemShut {NoStop}%
\bibitem [{\citenamefont {Tsui}\ \emph {et~al.}(1982)\citenamefont {Tsui},
  \citenamefont {Stormer},\ and\ \citenamefont {Gossard}}]{tsui1982two}%
  \BibitemOpen
  \bibfield  {author} {\bibinfo {author} {\bibfnamefont {D.~C.}\ \bibnamefont
  {Tsui}}, \bibinfo {author} {\bibfnamefont {H.~L.}\ \bibnamefont {Stormer}}, \
  and\ \bibinfo {author} {\bibfnamefont {A.~C.}\ \bibnamefont {Gossard}},\
  }\href {\doibase 10.1103/PhysRevLett.48.1559} {\bibfield  {journal} {\bibinfo
   {journal} {Phys. Rev. Lett.}\ }\textbf {\bibinfo {volume} {48}},\ \bibinfo
  {pages} {1559} (\bibinfo {year} {1982})}\BibitemShut {NoStop}%
\bibitem [{\citenamefont {Laughlin}(1981)}]{laughlin1981quantized}%
  \BibitemOpen
  \bibfield  {author} {\bibinfo {author} {\bibfnamefont {R.~B.}\ \bibnamefont
  {Laughlin}},\ }\href {\doibase 10.1103/PhysRevB.23.5632} {\bibfield
  {journal} {\bibinfo  {journal} {Phys. Rev. B}\ }\textbf {\bibinfo {volume}
  {23}},\ \bibinfo {pages} {5632} (\bibinfo {year} {1981})}\BibitemShut
  {NoStop}%
\bibitem [{\citenamefont {Laughlin}(1983)}]{laughlin1983anomalous}%
  \BibitemOpen
  \bibfield  {author} {\bibinfo {author} {\bibfnamefont {R.~B.}\ \bibnamefont
  {Laughlin}},\ }\href {\doibase 10.1103/PhysRevLett.50.1395} {\bibfield
  {journal} {\bibinfo  {journal} {Phys. Rev. Lett.}\ }\textbf {\bibinfo
  {volume} {50}},\ \bibinfo {pages} {1395} (\bibinfo {year}
  {1983})}\BibitemShut {NoStop}%
\bibitem [{\citenamefont {Maldacena}(1998)}]{maldacena1997the}%
  \BibitemOpen
  \bibfield  {author} {\bibinfo {author} {\bibfnamefont {J.~M.}\ \bibnamefont
  {Maldacena}},\ }\href {\doibase 10.4310/ATMP.1998.v2.n2.a1} {\bibfield
  {journal} {\bibinfo  {journal} {Adv. Theor. Math. Phys.}\ }\textbf {\bibinfo
  {volume} {2}},\ \bibinfo {pages} {231} (\bibinfo {year} {1998})},\ \Eprint
  {http://arxiv.org/abs/hep-th/9711200} {arXiv:hep-th/9711200} \BibitemShut
  {NoStop}%
\bibitem [{\citenamefont {Witten}(1998)}]{Witten1998anti}%
  \BibitemOpen
  \bibfield  {author} {\bibinfo {author} {\bibfnamefont {E.}~\bibnamefont
  {Witten}},\ }\href {\doibase 10.4310/ATMP.1998.v2.n2.a2} {\bibfield
  {journal} {\bibinfo  {journal} {Adv. Theor. Math. Phys.}\ }\textbf {\bibinfo
  {volume} {2}},\ \bibinfo {pages} {253} (\bibinfo {year} {1998})},\ \Eprint
  {http://arxiv.org/abs/hep-th/9802150} {arXiv:hep-th/9802150} \BibitemShut
  {NoStop}%
\bibitem [{\citenamefont {Gubser}\ \emph {et~al.}(1998)\citenamefont {Gubser},
  \citenamefont {Klebanov},\ and\ \citenamefont {Polyakov}}]{gubser1998gauge}%
  \BibitemOpen
  \bibfield  {author} {\bibinfo {author} {\bibfnamefont {S.~S.}\ \bibnamefont
  {Gubser}}, \bibinfo {author} {\bibfnamefont {I.~R.}\ \bibnamefont
  {Klebanov}}, \ and\ \bibinfo {author} {\bibfnamefont {A.~M.}\ \bibnamefont
  {Polyakov}},\ }\href {\doibase 10.1016/S0370-2693(98)00377-3} {\bibfield
  {journal} {\bibinfo  {journal} {Phys. Lett. B}\ }\textbf {\bibinfo {volume}
  {428}},\ \bibinfo {pages} {105} (\bibinfo {year} {1998})},\ \Eprint
  {http://arxiv.org/abs/hep-th/9802109} {arXiv:hep-th/9802109} \BibitemShut
  {NoStop}%
\bibitem [{\citenamefont {Polchinski}(1996)}]{polchinski1996TASI}%
  \BibitemOpen
  \bibfield  {author} {\bibinfo {author} {\bibfnamefont {J.}~\bibnamefont
  {Polchinski}},\ }in\ \href@noop {} {\emph {\bibinfo {booktitle} {{Theoretical
  Advanced Study Institute in Elementary Particle Physics (TASI 96): Fields,
  Strings, and Duality}}}}\ (\bibinfo {year} {1996})\ pp.\ \bibinfo {pages}
  {293--356},\ \Eprint {http://arxiv.org/abs/hep-th/9611050}
  {arXiv:hep-th/9611050} \BibitemShut {NoStop}%
\bibitem [{\citenamefont {Takayanagi}(2011)}]{takayanagi2011holographic}%
  \BibitemOpen
  \bibfield  {author} {\bibinfo {author} {\bibfnamefont {T.}~\bibnamefont
  {Takayanagi}},\ }\href {\doibase 10.1103/PhysRevLett.107.101602} {\bibfield
  {journal} {\bibinfo  {journal} {Phys. Rev. Lett.}\ }\textbf {\bibinfo
  {volume} {107}},\ \bibinfo {pages} {101602} (\bibinfo {year} {2011})},\
  \Eprint {http://arxiv.org/abs/1105.5165} {arXiv:1105.5165 [hep-th]}
  \BibitemShut {NoStop}%
\bibitem [{\citenamefont {Fujita}\ \emph {et~al.}(2011)\citenamefont {Fujita},
  \citenamefont {Takayanagi},\ and\ \citenamefont {Tonni}}]{Fujita_2011}%
  \BibitemOpen
  \bibfield  {author} {\bibinfo {author} {\bibfnamefont {M.}~\bibnamefont
  {Fujita}}, \bibinfo {author} {\bibfnamefont {T.}~\bibnamefont {Takayanagi}},
  \ and\ \bibinfo {author} {\bibfnamefont {E.}~\bibnamefont {Tonni}},\ }\href
  {\doibase 10.1007/jhep11(2011)043} {\bibfield  {journal} {\bibinfo  {journal}
  {Journal of High Energy Physics}\ }\textbf {\bibinfo {volume} {2011}}
  (\bibinfo {year} {2011}),\ 10.1007/jhep11(2011)043}\BibitemShut {NoStop}%
\bibitem [{\citenamefont {Recknagel}\ and\ \citenamefont
  {Schomerus}(2013)}]{Recknagel:2013uja}%
  \BibitemOpen
  \bibfield  {author} {\bibinfo {author} {\bibfnamefont {A.}~\bibnamefont
  {Recknagel}}\ and\ \bibinfo {author} {\bibfnamefont {V.}~\bibnamefont
  {Schomerus}},\ }\href {\doibase 10.1017/CBO9780511806476} {\emph {\bibinfo
  {title} {{Boundary Conformal Field Theory and the Worldsheet Approach to
  D-Branes}}}},\ Cambridge Monographs on Mathematical Physics\ (\bibinfo
  {publisher} {Cambridge University Press},\ \bibinfo {year}
  {2013})\BibitemShut {NoStop}%
\bibitem [{\citenamefont {Izumi}\ \emph {et~al.}(2022)\citenamefont {Izumi},
  \citenamefont {Shiromizu}, \citenamefont {Suzuki}, \citenamefont
  {Takayanagi},\ and\ \citenamefont {Tanahashi}}]{Izumi_2022}%
  \BibitemOpen
  \bibfield  {author} {\bibinfo {author} {\bibfnamefont {K.}~\bibnamefont
  {Izumi}}, \bibinfo {author} {\bibfnamefont {T.}~\bibnamefont {Shiromizu}},
  \bibinfo {author} {\bibfnamefont {K.}~\bibnamefont {Suzuki}}, \bibinfo
  {author} {\bibfnamefont {T.}~\bibnamefont {Takayanagi}}, \ and\ \bibinfo
  {author} {\bibfnamefont {N.}~\bibnamefont {Tanahashi}},\ }\href {\doibase
  10.1007/jhep10(2022)050} {\bibfield  {journal} {\bibinfo  {journal} {Journal
  of High Energy Physics}\ }\textbf {\bibinfo {volume} {2022}} (\bibinfo {year}
  {2022}),\ 10.1007/jhep10(2022)050}\BibitemShut {NoStop}%
\bibitem [{\citenamefont {Cardy}(1984)}]{cardy1984conformal}%
  \BibitemOpen
  \bibfield  {author} {\bibinfo {author} {\bibfnamefont {J.~L.}\ \bibnamefont
  {Cardy}},\ }\href {\doibase 10.1016/0550-3213(84)90241-4} {\bibfield
  {journal} {\bibinfo  {journal} {Nucl. Phys. B}\ }\textbf {\bibinfo {volume}
  {240}},\ \bibinfo {pages} {514} (\bibinfo {year} {1984})}\BibitemShut
  {NoStop}%
\bibitem [{\citenamefont {Cardy}(1989)}]{cardy1989boundary}%
  \BibitemOpen
  \bibfield  {author} {\bibinfo {author} {\bibfnamefont {J.~L.}\ \bibnamefont
  {Cardy}},\ }\href {\doibase 10.1016/0550-3213(89)90521-X} {\bibfield
  {journal} {\bibinfo  {journal} {Nucl. Phys. B}\ }\textbf {\bibinfo {volume}
  {324}},\ \bibinfo {pages} {581} (\bibinfo {year} {1989})}\BibitemShut
  {NoStop}%
\bibitem [{\citenamefont {Affleck}\ and\ \citenamefont
  {Ludwig}(1991)}]{affleck1991universal}%
  \BibitemOpen
  \bibfield  {author} {\bibinfo {author} {\bibfnamefont {I.}~\bibnamefont
  {Affleck}}\ and\ \bibinfo {author} {\bibfnamefont {A.~W.~W.}\ \bibnamefont
  {Ludwig}},\ }\href {\doibase 10.1103/PhysRevLett.67.161} {\bibfield
  {journal} {\bibinfo  {journal} {Phys. Rev. Lett.}\ }\textbf {\bibinfo
  {volume} {67}},\ \bibinfo {pages} {161} (\bibinfo {year} {1991})}\BibitemShut
  {NoStop}%
\bibitem [{\citenamefont
  {Cardy}(2008)}]{cardy2008boundaryconformalfieldtheory}%
  \BibitemOpen
  \bibfield  {author} {\bibinfo {author} {\bibfnamefont {J.}~\bibnamefont
  {Cardy}},\ }\href {https://arxiv.org/abs/hep-th/0411189} {\enquote {\bibinfo
  {title} {Boundary conformal field theory},}\ } (\bibinfo {year} {2008}),\
  \Eprint {http://arxiv.org/abs/hep-th/0411189} {arXiv:hep-th/0411189 [hep-th]}
  \BibitemShut {NoStop}%
\bibitem [{\citenamefont {Ge}\ and\ \citenamefont {Jian}(2024)}]{Ge:2024ubs}%
  \BibitemOpen
  \bibfield  {author} {\bibinfo {author} {\bibfnamefont {Y.}~\bibnamefont
  {Ge}}\ and\ \bibinfo {author} {\bibfnamefont {S.-K.}\ \bibnamefont {Jian}},\
  }\href@noop {} {\  (\bibinfo {year} {2024})},\ \Eprint
  {http://arxiv.org/abs/2403.18691} {arXiv:2403.18691 [cond-mat.stat-mech]}
  \BibitemShut {NoStop}%
\bibitem [{\citenamefont {Shen}(2023)}]{Shen:2023srk}%
  \BibitemOpen
  \bibfield  {author} {\bibinfo {author} {\bibfnamefont {X.-Y.}\ \bibnamefont
  {Shen}},\ }\href@noop {} {\  (\bibinfo {year} {2023})},\ \Eprint
  {http://arxiv.org/abs/2308.12598} {arXiv:2308.12598 [hep-th]} \BibitemShut
  {NoStop}%
\bibitem [{\citenamefont {Andrei}\ \emph {et~al.}(2018)\citenamefont {Andrei},
  \citenamefont {Bissi}, \citenamefont {Buican}, \citenamefont {Cardy},
  \citenamefont {Dorey}, \citenamefont {Drukker}, \citenamefont {Erdmenger},
  \citenamefont {Friedan}, \citenamefont {Fursaev}, \citenamefont {Konechny},
  \citenamefont {Kristjansen}, \citenamefont {Makabe}, \citenamefont
  {Nakayama}, \citenamefont {O'Bannon}, \citenamefont {Parini}, \citenamefont
  {Robinson}, \citenamefont {Ryu}, \citenamefont {Schmidt-Colinet},
  \citenamefont {Schomerus}, \citenamefont {Schweigert},\ and\ \citenamefont
  {Watts}}]{andrei2018boundarydefectcftopen}%
  \BibitemOpen
  \bibfield  {author} {\bibinfo {author} {\bibfnamefont {N.}~\bibnamefont
  {Andrei}}, \bibinfo {author} {\bibfnamefont {A.}~\bibnamefont {Bissi}},
  \bibinfo {author} {\bibfnamefont {M.}~\bibnamefont {Buican}}, \bibinfo
  {author} {\bibfnamefont {J.}~\bibnamefont {Cardy}}, \bibinfo {author}
  {\bibfnamefont {P.}~\bibnamefont {Dorey}}, \bibinfo {author} {\bibfnamefont
  {N.}~\bibnamefont {Drukker}}, \bibinfo {author} {\bibfnamefont
  {J.}~\bibnamefont {Erdmenger}}, \bibinfo {author} {\bibfnamefont
  {D.}~\bibnamefont {Friedan}}, \bibinfo {author} {\bibfnamefont
  {D.}~\bibnamefont {Fursaev}}, \bibinfo {author} {\bibfnamefont
  {A.}~\bibnamefont {Konechny}}, \bibinfo {author} {\bibfnamefont
  {C.}~\bibnamefont {Kristjansen}}, \bibinfo {author} {\bibfnamefont
  {I.}~\bibnamefont {Makabe}}, \bibinfo {author} {\bibfnamefont
  {Y.}~\bibnamefont {Nakayama}}, \bibinfo {author} {\bibfnamefont
  {A.}~\bibnamefont {O'Bannon}}, \bibinfo {author} {\bibfnamefont
  {R.}~\bibnamefont {Parini}}, \bibinfo {author} {\bibfnamefont
  {B.}~\bibnamefont {Robinson}}, \bibinfo {author} {\bibfnamefont
  {S.}~\bibnamefont {Ryu}}, \bibinfo {author} {\bibfnamefont {C.}~\bibnamefont
  {Schmidt-Colinet}}, \bibinfo {author} {\bibfnamefont {V.}~\bibnamefont
  {Schomerus}}, \bibinfo {author} {\bibfnamefont {C.}~\bibnamefont
  {Schweigert}}, \ and\ \bibinfo {author} {\bibfnamefont {G.}~\bibnamefont
  {Watts}},\ }\href {https://arxiv.org/abs/1810.05697} {\enquote {\bibinfo
  {title} {Boundary and defect cft: Open problems and applications},}\ }
  (\bibinfo {year} {2018}),\ \Eprint {http://arxiv.org/abs/1810.05697}
  {arXiv:1810.05697 [hep-th]} \BibitemShut {NoStop}%
\bibitem [{\citenamefont {Ji}\ and\ \citenamefont
  {Chen}(2024)}]{ji2024topologicaldefects21dsystems}%
  \BibitemOpen
  \bibfield  {author} {\bibinfo {author} {\bibfnamefont {W.}~\bibnamefont
  {Ji}}\ and\ \bibinfo {author} {\bibfnamefont {X.}~\bibnamefont {Chen}},\
  }\href {https://arxiv.org/abs/2407.02488} {\enquote {\bibinfo {title}
  {Topological defects of 2+1d systems from line excitations in 3+1d bulk},}\ }
  (\bibinfo {year} {2024}),\ \Eprint {http://arxiv.org/abs/2407.02488}
  {arXiv:2407.02488 [cond-mat.str-el]} \BibitemShut {NoStop}%
\bibitem [{\citenamefont {Brillaux}\ and\ \citenamefont
  {Fedorenko}(2021)}]{brillaux2021fermi}%
  \BibitemOpen
  \bibfield  {author} {\bibinfo {author} {\bibfnamefont {E.}~\bibnamefont
  {Brillaux}}\ and\ \bibinfo {author} {\bibfnamefont {A.~A.}\ \bibnamefont
  {Fedorenko}},\ }\href {\doibase 10.1103/PhysRevB.103.L081405} {\bibfield
  {journal} {\bibinfo  {journal} {Phys. Rev. B}\ }\textbf {\bibinfo {volume}
  {103}},\ \bibinfo {pages} {L081405} (\bibinfo {year} {2021})}\BibitemShut
  {NoStop}%
\bibitem [{\citenamefont {Brillaux}\ \emph {et~al.}(2024)\citenamefont
  {Brillaux}, \citenamefont {Fedorenko},\ and\ \citenamefont
  {Gruzberg}}]{brillaux2024surface}%
  \BibitemOpen
  \bibfield  {author} {\bibinfo {author} {\bibfnamefont {E.}~\bibnamefont
  {Brillaux}}, \bibinfo {author} {\bibfnamefont {A.~A.}\ \bibnamefont
  {Fedorenko}}, \ and\ \bibinfo {author} {\bibfnamefont {I.~A.}\ \bibnamefont
  {Gruzberg}},\ }\href {\doibase 10.1103/PhysRevB.109.174204} {\bibfield
  {journal} {\bibinfo  {journal} {Phys. Rev. B}\ }\textbf {\bibinfo {volume}
  {109}},\ \bibinfo {pages} {174204} (\bibinfo {year} {2024})}\BibitemShut
  {NoStop}%
\bibitem [{\citenamefont {Myerson-Jain}\ \emph {et~al.}(2024)\citenamefont
  {Myerson-Jain}, \citenamefont {Wu},\ and\ \citenamefont
  {Xu}}]{myerson2024pristine}%
  \BibitemOpen
  \bibfield  {author} {\bibinfo {author} {\bibfnamefont {N.}~\bibnamefont
  {Myerson-Jain}}, \bibinfo {author} {\bibfnamefont {X.-C.}\ \bibnamefont
  {Wu}}, \ and\ \bibinfo {author} {\bibfnamefont {C.}~\bibnamefont {Xu}},\
  }\href@noop {} {\bibfield  {journal} {\bibinfo  {journal} {arXiv preprint
  arXiv:2405.18481}\ } (\bibinfo {year} {2024})}\BibitemShut {NoStop}%
\bibitem [{\citenamefont {Ma}\ \emph {et~al.}(2022)\citenamefont {Ma},
  \citenamefont {Zou},\ and\ \citenamefont {Wang}}]{ma2022edge}%
  \BibitemOpen
  \bibfield  {author} {\bibinfo {author} {\bibfnamefont {R.}~\bibnamefont
  {Ma}}, \bibinfo {author} {\bibfnamefont {L.}~\bibnamefont {Zou}}, \ and\
  \bibinfo {author} {\bibfnamefont {C.}~\bibnamefont {Wang}},\ }\href {\doibase
  10.21468/SciPostPhys.12.6.196} {\bibfield  {journal} {\bibinfo  {journal}
  {SciPost Phys.}\ }\textbf {\bibinfo {volume} {12}},\ \bibinfo {pages} {196}
  (\bibinfo {year} {2022})}\BibitemShut {NoStop}%
\bibitem [{\citenamefont {Kitaev}(2009)}]{kitaev2009periodic}%
  \BibitemOpen
  \bibfield  {author} {\bibinfo {author} {\bibfnamefont {A.}~\bibnamefont
  {Kitaev}},\ }\href {\doibase 10.1063/1.3149495} {\bibfield  {journal}
  {\bibinfo  {journal} {AIP Conference Proceedings}\ }\textbf {\bibinfo
  {volume} {1134}},\ \bibinfo {pages} {22} (\bibinfo {year} {2009})},\ \Eprint
  {http://arxiv.org/abs/https://pubs.aip.org/aip/acp/article-pdf/1134/1/22/11584243/22\_1\_online.pdf}
  {https://pubs.aip.org/aip/acp/article-pdf/1134/1/22/11584243/22\_1\_online.pdf}
  \BibitemShut {NoStop}%
\bibitem [{\citenamefont {Ryu}\ \emph {et~al.}(2010)\citenamefont {Ryu},
  \citenamefont {Schnyder}, \citenamefont {Furusaki},\ and\ \citenamefont
  {Ludwig}}]{ryu2010topological}%
  \BibitemOpen
  \bibfield  {author} {\bibinfo {author} {\bibfnamefont {S.}~\bibnamefont
  {Ryu}}, \bibinfo {author} {\bibfnamefont {A.~P.}\ \bibnamefont {Schnyder}},
  \bibinfo {author} {\bibfnamefont {A.}~\bibnamefont {Furusaki}}, \ and\
  \bibinfo {author} {\bibfnamefont {A.~W.}\ \bibnamefont {Ludwig}},\
  }\href@noop {} {\bibfield  {journal} {\bibinfo  {journal} {New Journal of
  Physics}\ }\textbf {\bibinfo {volume} {12}},\ \bibinfo {pages} {065010}
  (\bibinfo {year} {2010})}\BibitemShut {NoStop}%
\bibitem [{\citenamefont {Gu}\ and\ \citenamefont {Wen}(2009)}]{Gu_2009}%
  \BibitemOpen
  \bibfield  {author} {\bibinfo {author} {\bibfnamefont {Z.-C.}\ \bibnamefont
  {Gu}}\ and\ \bibinfo {author} {\bibfnamefont {X.-G.}\ \bibnamefont {Wen}},\
  }\href {\doibase 10.1103/physrevb.80.155131} {\bibfield  {journal} {\bibinfo
  {journal} {Physical Review B}\ }\textbf {\bibinfo {volume} {80}} (\bibinfo
  {year} {2009}),\ 10.1103/physrevb.80.155131}\BibitemShut {NoStop}%
\bibitem [{\citenamefont {Hasan}\ and\ \citenamefont
  {Kane}(2010)}]{hasan2010toopological}%
  \BibitemOpen
  \bibfield  {author} {\bibinfo {author} {\bibfnamefont {M.~Z.}\ \bibnamefont
  {Hasan}}\ and\ \bibinfo {author} {\bibfnamefont {C.~L.}\ \bibnamefont
  {Kane}},\ }\href {\doibase 10.1103/RevModPhys.82.3045} {\bibfield  {journal}
  {\bibinfo  {journal} {Rev. Mod. Phys.}\ }\textbf {\bibinfo {volume} {82}},\
  \bibinfo {pages} {3045} (\bibinfo {year} {2010})}\BibitemShut {NoStop}%
\bibitem [{\citenamefont {Qi}\ and\ \citenamefont {Zhang}(2011)}]{Qi_2011}%
  \BibitemOpen
  \bibfield  {author} {\bibinfo {author} {\bibfnamefont {X.-L.}\ \bibnamefont
  {Qi}}\ and\ \bibinfo {author} {\bibfnamefont {S.-C.}\ \bibnamefont {Zhang}},\
  }\href {\doibase 10.1103/revmodphys.83.1057} {\bibfield  {journal} {\bibinfo
  {journal} {Reviews of Modern Physics}\ }\textbf {\bibinfo {volume} {83}},\
  \bibinfo {pages} {1057–1110} (\bibinfo {year} {2011})}\BibitemShut
  {NoStop}%
\bibitem [{\citenamefont {Chen}\ \emph {et~al.}(2012)\citenamefont {Chen},
  \citenamefont {Gu}, \citenamefont {Liu},\ and\ \citenamefont
  {Wen}}]{chen2012symmetry}%
  \BibitemOpen
  \bibfield  {author} {\bibinfo {author} {\bibfnamefont {X.}~\bibnamefont
  {Chen}}, \bibinfo {author} {\bibfnamefont {Z.-C.}\ \bibnamefont {Gu}},
  \bibinfo {author} {\bibfnamefont {Z.-X.}\ \bibnamefont {Liu}}, \ and\
  \bibinfo {author} {\bibfnamefont {X.-G.}\ \bibnamefont {Wen}},\ }\href
  {\doibase 10.1126/science.1227224} {\bibfield  {journal} {\bibinfo  {journal}
  {Science}\ }\textbf {\bibinfo {volume} {338}},\ \bibinfo {pages} {1604}
  (\bibinfo {year} {2012})},\ \Eprint
  {http://arxiv.org/abs/https://www.science.org/doi/pdf/10.1126/science.1227224}
  {https://www.science.org/doi/pdf/10.1126/science.1227224} \BibitemShut
  {NoStop}%
\bibitem [{\citenamefont {Kane}\ and\ \citenamefont
  {Mele}(2005)}]{kane2005quantum}%
  \BibitemOpen
  \bibfield  {author} {\bibinfo {author} {\bibfnamefont {C.~L.}\ \bibnamefont
  {Kane}}\ and\ \bibinfo {author} {\bibfnamefont {E.~J.}\ \bibnamefont
  {Mele}},\ }\href {\doibase 10.1103/PhysRevLett.95.226801} {\bibfield
  {journal} {\bibinfo  {journal} {Phys. Rev. Lett.}\ }\textbf {\bibinfo
  {volume} {95}},\ \bibinfo {pages} {226801} (\bibinfo {year}
  {2005})}\BibitemShut {NoStop}%
\bibitem [{\citenamefont {Bernevig}\ \emph {et~al.}(2006)\citenamefont
  {Bernevig}, \citenamefont {Hughes},\ and\ \citenamefont
  {Zhang}}]{bernevig2006quantum}%
  \BibitemOpen
  \bibfield  {author} {\bibinfo {author} {\bibfnamefont {B.~A.}\ \bibnamefont
  {Bernevig}}, \bibinfo {author} {\bibfnamefont {T.~L.}\ \bibnamefont
  {Hughes}}, \ and\ \bibinfo {author} {\bibfnamefont {S.-C.}\ \bibnamefont
  {Zhang}},\ }\href {\doibase 10.1126/science.1133734} {\bibfield  {journal}
  {\bibinfo  {journal} {Science}\ }\textbf {\bibinfo {volume} {314}},\ \bibinfo
  {pages} {1757} (\bibinfo {year} {2006})},\ \Eprint
  {http://arxiv.org/abs/https://www.science.org/doi/pdf/10.1126/science.1133734}
  {https://www.science.org/doi/pdf/10.1126/science.1133734} \BibitemShut
  {NoStop}%
\bibitem [{\citenamefont {König}\ \emph {et~al.}(2007)\citenamefont {König},
  \citenamefont {Wiedmann}, \citenamefont {Brüne}, \citenamefont {Roth},
  \citenamefont {Buhmann}, \citenamefont {Molenkamp}, \citenamefont {Qi},\ and\
  \citenamefont {Zhang}}]{molenkamp2007quantum}%
  \BibitemOpen
  \bibfield  {author} {\bibinfo {author} {\bibfnamefont {M.}~\bibnamefont
  {König}}, \bibinfo {author} {\bibfnamefont {S.}~\bibnamefont {Wiedmann}},
  \bibinfo {author} {\bibfnamefont {C.}~\bibnamefont {Brüne}}, \bibinfo
  {author} {\bibfnamefont {A.}~\bibnamefont {Roth}}, \bibinfo {author}
  {\bibfnamefont {H.}~\bibnamefont {Buhmann}}, \bibinfo {author} {\bibfnamefont
  {L.~W.}\ \bibnamefont {Molenkamp}}, \bibinfo {author} {\bibfnamefont {X.-L.}\
  \bibnamefont {Qi}}, \ and\ \bibinfo {author} {\bibfnamefont {S.-C.}\
  \bibnamefont {Zhang}},\ }\href {\doibase 10.1126/science.1148047} {\bibfield
  {journal} {\bibinfo  {journal} {Science}\ }\textbf {\bibinfo {volume}
  {318}},\ \bibinfo {pages} {766} (\bibinfo {year} {2007})},\ \Eprint
  {http://arxiv.org/abs/https://www.science.org/doi/pdf/10.1126/science.1148047}
  {https://www.science.org/doi/pdf/10.1126/science.1148047} \BibitemShut
  {NoStop}%
\bibitem [{\citenamefont {Hsieh}\ \emph {et~al.}(2008)\citenamefont {Hsieh},
  \citenamefont {Qian}, \citenamefont {Wray}, \citenamefont {Xia},
  \citenamefont {Hor}, \citenamefont {Cava},\ and\ \citenamefont
  {Hasan}}]{hsieh2008topological}%
  \BibitemOpen
  \bibfield  {author} {\bibinfo {author} {\bibfnamefont {D.}~\bibnamefont
  {Hsieh}}, \bibinfo {author} {\bibfnamefont {D.}~\bibnamefont {Qian}},
  \bibinfo {author} {\bibfnamefont {L.}~\bibnamefont {Wray}}, \bibinfo {author}
  {\bibfnamefont {Y.}~\bibnamefont {Xia}}, \bibinfo {author} {\bibfnamefont
  {Y.~S.}\ \bibnamefont {Hor}}, \bibinfo {author} {\bibfnamefont {R.~J.}\
  \bibnamefont {Cava}}, \ and\ \bibinfo {author} {\bibfnamefont {M.~Z.}\
  \bibnamefont {Hasan}},\ }\href@noop {} {\bibfield  {journal} {\bibinfo
  {journal} {Nature}\ }\textbf {\bibinfo {volume} {452}},\ \bibinfo {pages}
  {970} (\bibinfo {year} {2008})}\BibitemShut {NoStop}%
\bibitem [{\citenamefont {Zhang}\ and\ \citenamefont
  {Wang}(2017)}]{zhang2017unconventional}%
  \BibitemOpen
  \bibfield  {author} {\bibinfo {author} {\bibfnamefont {L.}~\bibnamefont
  {Zhang}}\ and\ \bibinfo {author} {\bibfnamefont {F.}~\bibnamefont {Wang}},\
  }\href {\doibase 10.1103/PhysRevLett.118.087201} {\bibfield  {journal}
  {\bibinfo  {journal} {Phys. Rev. Lett.}\ }\textbf {\bibinfo {volume} {118}},\
  \bibinfo {pages} {087201} (\bibinfo {year} {2017})}\BibitemShut {NoStop}%
\bibitem [{\citenamefont {Wu}\ \emph {et~al.}(2020)\citenamefont {Wu},
  \citenamefont {Xu}, \citenamefont {Geng}, \citenamefont {Jian},\ and\
  \citenamefont {Xu}}]{wu2020boundary}%
  \BibitemOpen
  \bibfield  {author} {\bibinfo {author} {\bibfnamefont {X.-C.}\ \bibnamefont
  {Wu}}, \bibinfo {author} {\bibfnamefont {Y.}~\bibnamefont {Xu}}, \bibinfo
  {author} {\bibfnamefont {H.}~\bibnamefont {Geng}}, \bibinfo {author}
  {\bibfnamefont {C.-M.}\ \bibnamefont {Jian}}, \ and\ \bibinfo {author}
  {\bibfnamefont {C.}~\bibnamefont {Xu}},\ }\href {\doibase
  10.1103/PhysRevB.101.174406} {\bibfield  {journal} {\bibinfo  {journal}
  {Phys. Rev. B}\ }\textbf {\bibinfo {volume} {101}},\ \bibinfo {pages}
  {174406} (\bibinfo {year} {2020})}\BibitemShut {NoStop}%
\bibitem [{\citenamefont {Scaffidi}\ \emph {et~al.}(2017)\citenamefont
  {Scaffidi}, \citenamefont {Parker},\ and\ \citenamefont
  {Vasseur}}]{scaffidi2017gapless}%
  \BibitemOpen
  \bibfield  {author} {\bibinfo {author} {\bibfnamefont {T.}~\bibnamefont
  {Scaffidi}}, \bibinfo {author} {\bibfnamefont {D.~E.}\ \bibnamefont
  {Parker}}, \ and\ \bibinfo {author} {\bibfnamefont {R.}~\bibnamefont
  {Vasseur}},\ }\href {\doibase 10.1103/PhysRevX.7.041048} {\bibfield
  {journal} {\bibinfo  {journal} {Phys. Rev. X}\ }\textbf {\bibinfo {volume}
  {7}},\ \bibinfo {pages} {041048} (\bibinfo {year} {2017})}\BibitemShut
  {NoStop}%
\bibitem [{\citenamefont {Verresen}(2020)}]{verresen2020topology}%
  \BibitemOpen
  \bibfield  {author} {\bibinfo {author} {\bibfnamefont {R.}~\bibnamefont
  {Verresen}},\ }\href@noop {} {\bibfield  {journal} {\bibinfo  {journal}
  {arXiv preprint arXiv:2003.05453}\ } (\bibinfo {year} {2020})}\BibitemShut
  {NoStop}%
\bibitem [{\citenamefont {Verresen}\ \emph {et~al.}(2021)\citenamefont
  {Verresen}, \citenamefont {Thorngren}, \citenamefont {Jones},\ and\
  \citenamefont {Pollmann}}]{verresen2021gapless}%
  \BibitemOpen
  \bibfield  {author} {\bibinfo {author} {\bibfnamefont {R.}~\bibnamefont
  {Verresen}}, \bibinfo {author} {\bibfnamefont {R.}~\bibnamefont {Thorngren}},
  \bibinfo {author} {\bibfnamefont {N.~G.}\ \bibnamefont {Jones}}, \ and\
  \bibinfo {author} {\bibfnamefont {F.}~\bibnamefont {Pollmann}},\ }\href
  {\doibase 10.1103/PhysRevX.11.041059} {\bibfield  {journal} {\bibinfo
  {journal} {Phys. Rev. X}\ }\textbf {\bibinfo {volume} {11}},\ \bibinfo
  {pages} {041059} (\bibinfo {year} {2021})}\BibitemShut {NoStop}%
\bibitem [{\citenamefont {Yu}\ \emph {et~al.}(2022)\citenamefont {Yu},
  \citenamefont {Huang}, \citenamefont {Song}, \citenamefont {Xu},
  \citenamefont {Ding},\ and\ \citenamefont {Zhang}}]{yu2022conformal}%
  \BibitemOpen
  \bibfield  {author} {\bibinfo {author} {\bibfnamefont {X.-J.}\ \bibnamefont
  {Yu}}, \bibinfo {author} {\bibfnamefont {R.-Z.}\ \bibnamefont {Huang}},
  \bibinfo {author} {\bibfnamefont {H.-H.}\ \bibnamefont {Song}}, \bibinfo
  {author} {\bibfnamefont {L.}~\bibnamefont {Xu}}, \bibinfo {author}
  {\bibfnamefont {C.}~\bibnamefont {Ding}}, \ and\ \bibinfo {author}
  {\bibfnamefont {L.}~\bibnamefont {Zhang}},\ }\href {\doibase
  10.1103/PhysRevLett.129.210601} {\bibfield  {journal} {\bibinfo  {journal}
  {Phys. Rev. Lett.}\ }\textbf {\bibinfo {volume} {129}},\ \bibinfo {pages}
  {210601} (\bibinfo {year} {2022})}\BibitemShut {NoStop}%
\bibitem [{\citenamefont {Yu}\ \emph {et~al.}(2024)\citenamefont {Yu},
  \citenamefont {Yang}, \citenamefont {Lin},\ and\ \citenamefont
  {Jian}}]{yu2024universal}%
  \BibitemOpen
  \bibfield  {author} {\bibinfo {author} {\bibfnamefont {X.-J.}\ \bibnamefont
  {Yu}}, \bibinfo {author} {\bibfnamefont {S.}~\bibnamefont {Yang}}, \bibinfo
  {author} {\bibfnamefont {H.-Q.}\ \bibnamefont {Lin}}, \ and\ \bibinfo
  {author} {\bibfnamefont {S.-K.}\ \bibnamefont {Jian}},\ }\href {\doibase
  10.1103/PhysRevLett.133.026601} {\bibfield  {journal} {\bibinfo  {journal}
  {Phys. Rev. Lett.}\ }\textbf {\bibinfo {volume} {133}},\ \bibinfo {pages}
  {026601} (\bibinfo {year} {2024})}\BibitemShut {NoStop}%
\bibitem [{\citenamefont {Liu}\ \emph {et~al.}(2021)\citenamefont {Liu},
  \citenamefont {Shapourian}, \citenamefont {Vishwanath},\ and\ \citenamefont
  {Metlitski}}]{Liu:2021nck}%
  \BibitemOpen
  \bibfield  {author} {\bibinfo {author} {\bibfnamefont {S.}~\bibnamefont
  {Liu}}, \bibinfo {author} {\bibfnamefont {H.}~\bibnamefont {Shapourian}},
  \bibinfo {author} {\bibfnamefont {A.}~\bibnamefont {Vishwanath}}, \ and\
  \bibinfo {author} {\bibfnamefont {M.~A.}\ \bibnamefont {Metlitski}},\ }\href
  {\doibase 10.1103/PhysRevB.104.104201} {\bibfield  {journal} {\bibinfo
  {journal} {Phys. Rev. B}\ }\textbf {\bibinfo {volume} {104}},\ \bibinfo
  {pages} {104201} (\bibinfo {year} {2021})},\ \Eprint
  {http://arxiv.org/abs/2104.15026} {arXiv:2104.15026 [cond-mat.str-el]}
  \BibitemShut {NoStop}%
\bibitem [{\citenamefont {{Giombi}}\ \emph {et~al.}(2022)\citenamefont
  {{Giombi}}, \citenamefont {{Helfenberger}},\ and\ \citenamefont
  {{Khanchandani}}}]{2022Giombia}%
  \BibitemOpen
  \bibfield  {author} {\bibinfo {author} {\bibfnamefont {S.}~\bibnamefont
  {{Giombi}}}, \bibinfo {author} {\bibfnamefont {E.}~\bibnamefont
  {{Helfenberger}}}, \ and\ \bibinfo {author} {\bibfnamefont {H.}~\bibnamefont
  {{Khanchandani}}},\ }\href {\doibase 10.1007/JHEP07(2022)018} {\bibfield
  {journal} {\bibinfo  {journal} {Journal of High Energy Physics}\ }\textbf
  {\bibinfo {volume} {2022}},\ \bibinfo {eid} {18} (\bibinfo {year} {2022})},\
  \Eprint {http://arxiv.org/abs/2110.04268} {arXiv:2110.04268 [hep-th]}
  \BibitemShut {NoStop}%
\bibitem [{\citenamefont {{Giombi}}\ \emph {et~al.}(2023)\citenamefont
  {{Giombi}}, \citenamefont {{Helfenberger}},\ and\ \citenamefont
  {{Khanchandani}}}]{2023Giombib}%
  \BibitemOpen
  \bibfield  {author} {\bibinfo {author} {\bibfnamefont {S.}~\bibnamefont
  {{Giombi}}}, \bibinfo {author} {\bibfnamefont {E.}~\bibnamefont
  {{Helfenberger}}}, \ and\ \bibinfo {author} {\bibfnamefont {H.}~\bibnamefont
  {{Khanchandani}}},\ }\href {\doibase 10.1007/JHEP08(2023)224} {\bibfield
  {journal} {\bibinfo  {journal} {Journal of High Energy Physics}\ }\textbf
  {\bibinfo {volume} {2023}},\ \bibinfo {eid} {224} (\bibinfo {year} {2023})},\
  \Eprint {http://arxiv.org/abs/2211.11073} {arXiv:2211.11073 [hep-th]}
  \BibitemShut {NoStop}%
\bibitem [{\citenamefont {{Barrat}}\ \emph {et~al.}(2023)\citenamefont
  {{Barrat}}, \citenamefont {{Liendo}},\ and\ \citenamefont {{van
  Vliet}}}]{2023Barrat}%
  \BibitemOpen
  \bibfield  {author} {\bibinfo {author} {\bibfnamefont {J.}~\bibnamefont
  {{Barrat}}}, \bibinfo {author} {\bibfnamefont {P.}~\bibnamefont {{Liendo}}},
  \ and\ \bibinfo {author} {\bibfnamefont {P.}~\bibnamefont {{van Vliet}}},\
  }\href {\doibase 10.48550/arXiv.2304.13588} {\bibfield  {journal} {\bibinfo
  {journal} {arXiv e-prints}\ ,\ \bibinfo {eid} {arXiv:2304.13588}} (\bibinfo
  {year} {2023})},\ \Eprint {http://arxiv.org/abs/2304.13588} {arXiv:2304.13588
  [hep-th]} \BibitemShut {NoStop}%
\bibitem [{\citenamefont {Shachar}\ \emph {et~al.}(2024)\citenamefont
  {Shachar}, \citenamefont {Sinha},\ and\ \citenamefont
  {Smolkin}}]{Shachar:2024ubf}%
  \BibitemOpen
  \bibfield  {author} {\bibinfo {author} {\bibfnamefont {T.}~\bibnamefont
  {Shachar}}, \bibinfo {author} {\bibfnamefont {R.}~\bibnamefont {Sinha}}, \
  and\ \bibinfo {author} {\bibfnamefont {M.}~\bibnamefont {Smolkin}},\
  }\href@noop {} {\  (\bibinfo {year} {2024})},\ \Eprint
  {http://arxiv.org/abs/2404.18403} {arXiv:2404.18403 [hep-th]} \BibitemShut
  {NoStop}%
\bibitem [{\citenamefont {Herzog}\ and\ \citenamefont
  {Schaub}(2023)}]{Herzog_2023}%
  \BibitemOpen
  \bibfield  {author} {\bibinfo {author} {\bibfnamefont {C.~P.}\ \bibnamefont
  {Herzog}}\ and\ \bibinfo {author} {\bibfnamefont {V.}~\bibnamefont
  {Schaub}},\ }\href {\doibase 10.1007/jhep02(2023)129} {\bibfield  {journal}
  {\bibinfo  {journal} {Journal of High Energy Physics}\ }\textbf {\bibinfo
  {volume} {2023}} (\bibinfo {year} {2023}),\
  10.1007/jhep02(2023)129}\BibitemShut {NoStop}%
\bibitem [{\citenamefont {{Diehl}}\ and\ \citenamefont
  {{Dietrich}}(1981)}]{1981Diehl}%
  \BibitemOpen
  \bibfield  {author} {\bibinfo {author} {\bibfnamefont {H.~W.}\ \bibnamefont
  {{Diehl}}}\ and\ \bibinfo {author} {\bibfnamefont {S.}~\bibnamefont
  {{Dietrich}}},\ }\href {\doibase 10.1007/BF01298293} {\bibfield  {journal}
  {\bibinfo  {journal} {Zeitschrift fur Physik B Condensed Matter}\ }\textbf
  {\bibinfo {volume} {42}},\ \bibinfo {pages} {65} (\bibinfo {year}
  {1981})}\BibitemShut {NoStop}%
\bibitem [{\citenamefont {{Diehl}}\ and\ \citenamefont
  {{Dietrich}}(1983)}]{1983Diehl}%
  \BibitemOpen
  \bibfield  {author} {\bibinfo {author} {\bibfnamefont {H.~W.}\ \bibnamefont
  {{Diehl}}}\ and\ \bibinfo {author} {\bibfnamefont {S.}~\bibnamefont
  {{Dietrich}}},\ }\href {\doibase 10.1007/BF01304094} {\bibfield  {journal}
  {\bibinfo  {journal} {Zeitschrift fur Physik B Condensed Matter}\ }\textbf
  {\bibinfo {volume} {50}},\ \bibinfo {pages} {117} (\bibinfo {year}
  {1983})}\BibitemShut {NoStop}%
\bibitem [{\citenamefont {{Giombi}}\ and\ \citenamefont
  {{Khanchandani}}(2020)}]{2020Giombi}%
  \BibitemOpen
  \bibfield  {author} {\bibinfo {author} {\bibfnamefont {S.}~\bibnamefont
  {{Giombi}}}\ and\ \bibinfo {author} {\bibfnamefont {H.}~\bibnamefont
  {{Khanchandani}}},\ }\href {\doibase 10.1007/JHEP08(2020)010} {\bibfield
  {journal} {\bibinfo  {journal} {Journal of High Energy Physics}\ }\textbf
  {\bibinfo {volume} {2020}},\ \bibinfo {eid} {10} (\bibinfo {year} {2020})},\
  \Eprint {http://arxiv.org/abs/1912.08169} {arXiv:1912.08169 [hep-th]}
  \BibitemShut {NoStop}%
\bibitem [{\citenamefont {{Padayasi}}\ \emph {et~al.}(2022)\citenamefont
  {{Padayasi}}, \citenamefont {{Krishnan}}, \citenamefont {{Metlitski}},
  \citenamefont {{Gruzberg}},\ and\ \citenamefont {{Meineri}}}]{2022Padayasi}%
  \BibitemOpen
  \bibfield  {author} {\bibinfo {author} {\bibfnamefont {J.}~\bibnamefont
  {{Padayasi}}}, \bibinfo {author} {\bibfnamefont {A.}~\bibnamefont
  {{Krishnan}}}, \bibinfo {author} {\bibfnamefont {M.}~\bibnamefont
  {{Metlitski}}}, \bibinfo {author} {\bibfnamefont {I.}~\bibnamefont
  {{Gruzberg}}}, \ and\ \bibinfo {author} {\bibfnamefont {M.}~\bibnamefont
  {{Meineri}}},\ }\href {\doibase 10.21468/SciPostPhys.12.6.190} {\bibfield
  {journal} {\bibinfo  {journal} {SciPost Physics}\ }\textbf {\bibinfo {volume}
  {12}},\ \bibinfo {eid} {190} (\bibinfo {year} {2022})},\ \Eprint
  {http://arxiv.org/abs/2111.03071} {arXiv:2111.03071 [cond-mat.stat-mech]}
  \BibitemShut {NoStop}%
\bibitem [{\citenamefont {{Parisen Toldin}}\ and\ \citenamefont
  {{Metlitski}}(2022)}]{2022Parisen}%
  \BibitemOpen
  \bibfield  {author} {\bibinfo {author} {\bibfnamefont {F.}~\bibnamefont
  {{Parisen Toldin}}}\ and\ \bibinfo {author} {\bibfnamefont {M.~A.}\
  \bibnamefont {{Metlitski}}},\ }\href {\doibase
  10.1103/PhysRevLett.128.215701} {\bibfield  {journal} {\bibinfo  {journal}
  {\prl}\ }\textbf {\bibinfo {volume} {128}},\ \bibinfo {eid} {215701}
  (\bibinfo {year} {2022})},\ \Eprint {http://arxiv.org/abs/2111.03613}
  {arXiv:2111.03613 [cond-mat.stat-mech]} \BibitemShut {NoStop}%
\bibitem [{\citenamefont {{Metlitski}}(2022)}]{2022metlitski}%
  \BibitemOpen
  \bibfield  {author} {\bibinfo {author} {\bibfnamefont {M.}~\bibnamefont
  {{Metlitski}}},\ }\href {\doibase 10.21468/SciPostPhys.12.4.131} {\bibfield
  {journal} {\bibinfo  {journal} {SciPost Physics}\ }\textbf {\bibinfo {volume}
  {12}},\ \bibinfo {eid} {131} (\bibinfo {year} {2022})}\BibitemShut {NoStop}%
\bibitem [{\citenamefont {Kosterlitz}\ and\ \citenamefont
  {Thouless}(2018)}]{kosterlitz2018ordering}%
  \BibitemOpen
  \bibfield  {author} {\bibinfo {author} {\bibfnamefont {J.~M.}\ \bibnamefont
  {Kosterlitz}}\ and\ \bibinfo {author} {\bibfnamefont {D.~J.}\ \bibnamefont
  {Thouless}},\ }in\ \href@noop {} {\emph {\bibinfo {booktitle} {Basic Notions
  Of Condensed Matter Physics}}}\ (\bibinfo  {publisher} {CRC Press},\ \bibinfo
  {year} {2018})\ pp.\ \bibinfo {pages} {493--515}\BibitemShut {NoStop}%
\bibitem [{\citenamefont {Berezinsky}(1971)}]{Berezinsky:1970fr}%
  \BibitemOpen
  \bibfield  {author} {\bibinfo {author} {\bibfnamefont {V.~L.}\ \bibnamefont
  {Berezinsky}},\ }\href@noop {} {\bibfield  {journal} {\bibinfo  {journal}
  {Sov. Phys. JETP}\ }\textbf {\bibinfo {volume} {32}},\ \bibinfo {pages} {493}
  (\bibinfo {year} {1971})}\BibitemShut {NoStop}%
\bibitem [{\citenamefont {Chang}\ \emph {et~al.}(2023)\citenamefont {Chang},
  \citenamefont {Chen},\ and\ \citenamefont {Xu}}]{Chang:2022hud}%
  \BibitemOpen
  \bibfield  {author} {\bibinfo {author} {\bibfnamefont {C.-M.}\ \bibnamefont
  {Chang}}, \bibinfo {author} {\bibfnamefont {J.}~\bibnamefont {Chen}}, \ and\
  \bibinfo {author} {\bibfnamefont {F.}~\bibnamefont {Xu}},\ }\href {\doibase
  10.21468/SciPostPhys.15.5.216} {\bibfield  {journal} {\bibinfo  {journal}
  {SciPost Phys.}\ }\textbf {\bibinfo {volume} {15}},\ \bibinfo {pages} {216}
  (\bibinfo {year} {2023})},\ \Eprint {http://arxiv.org/abs/2208.02757}
  {arXiv:2208.02757 [hep-th]} \BibitemShut {NoStop}%
\bibitem [{\citenamefont {{Giombi}}\ and\ \citenamefont
  {{Liu}}(2023)}]{2023Simone}%
  \BibitemOpen
  \bibfield  {author} {\bibinfo {author} {\bibfnamefont {S.}~\bibnamefont
  {{Giombi}}}\ and\ \bibinfo {author} {\bibfnamefont {B.}~\bibnamefont
  {{Liu}}},\ }\href {\doibase 10.1007/JHEP12(2023)004} {\bibfield  {journal}
  {\bibinfo  {journal} {Journal of High Energy Physics}\ }\textbf {\bibinfo
  {volume} {2023}},\ \bibinfo {eid} {4} (\bibinfo {year} {2023})},\ \Eprint
  {http://arxiv.org/abs/2305.11402} {arXiv:2305.11402 [hep-th]} \BibitemShut
  {NoStop}%
\bibitem [{\citenamefont {Tr\'epanier}(2023)}]{Trepanier:2023tvb}%
  \BibitemOpen
  \bibfield  {author} {\bibinfo {author} {\bibfnamefont {M.}~\bibnamefont
  {Tr\'epanier}},\ }\href {\doibase 10.1007/JHEP09(2023)074} {\bibfield
  {journal} {\bibinfo  {journal} {JHEP}\ }\textbf {\bibinfo {volume} {09}},\
  \bibinfo {pages} {074} (\bibinfo {year} {2023})},\ \Eprint
  {http://arxiv.org/abs/2305.10486} {arXiv:2305.10486 [hep-th]} \BibitemShut
  {NoStop}%
\bibitem [{Note1()}]{Note1}%
  \BibitemOpen
  \bibinfo {note} {Some boundary symmetry-breaking coupling that drives the
  system to the normal universality class is neglected. See Diehl \cite
  {1981Diehl,1983Diehl} for a detailed discussion.}\BibitemShut {Stop}%
\bibitem [{Note2()}]{Note2}%
  \BibitemOpen
  \bibinfo {note} {Strictly speaking, the surface mass is located at some
  $h=h_\protect \text {sp}$ for the special transition.}\BibitemShut {Stop}%
\bibitem [{\citenamefont {{Herzog}}\ and\ \citenamefont
  {{Huang}}(2017)}]{2017Christopher}%
  \BibitemOpen
  \bibfield  {author} {\bibinfo {author} {\bibfnamefont {C.~P.}\ \bibnamefont
  {{Herzog}}}\ and\ \bibinfo {author} {\bibfnamefont {K.-W.}\ \bibnamefont
  {{Huang}}},\ }\href {\doibase 10.1007/JHEP10(2017)189} {\bibfield  {journal}
  {\bibinfo  {journal} {Journal of High Energy Physics}\ }\textbf {\bibinfo
  {volume} {2017}},\ \bibinfo {eid} {189} (\bibinfo {year} {2017})},\ \Eprint
  {http://arxiv.org/abs/1707.06224} {arXiv:1707.06224 [hep-th]} \BibitemShut
  {NoStop}%
\bibitem [{Note3()}]{Note3}%
  \BibitemOpen
  \bibinfo {note} {For clarity, we emphasize that the terminology ``boundary
  criticality'' in this paper refers to the criticality that originates from
  interplay between different dimensional degrees of freedom, i.e. gapless bulk
  bosons and gapless edge modes in the context of topological insulator and
  superconductor. The theory constructed solely via boundary degrees of freedom
  at the surface transition does not fall into this category.}\BibitemShut
  {Stop}%
\bibitem [{\citenamefont {Friedan}\ \emph {et~al.}(1985)\citenamefont
  {Friedan}, \citenamefont {Qiu},\ and\ \citenamefont
  {Shenker}}]{friedan1984superconformal}%
  \BibitemOpen
  \bibfield  {author} {\bibinfo {author} {\bibfnamefont {D.}~\bibnamefont
  {Friedan}}, \bibinfo {author} {\bibfnamefont {Z.-a.}\ \bibnamefont {Qiu}}, \
  and\ \bibinfo {author} {\bibfnamefont {S.~H.}\ \bibnamefont {Shenker}},\
  }\href {\doibase 10.1016/0370-2693(85)90819-6} {\bibfield  {journal}
  {\bibinfo  {journal} {Phys. Lett. B}\ }\textbf {\bibinfo {volume} {151}},\
  \bibinfo {pages} {37} (\bibinfo {year} {1985})}\BibitemShut {NoStop}%
\bibitem [{\citenamefont {Grover}\ \emph {et~al.}(2014)\citenamefont {Grover},
  \citenamefont {Sheng},\ and\ \citenamefont {Vishwanath}}]{Grover:2013rc}%
  \BibitemOpen
  \bibfield  {author} {\bibinfo {author} {\bibfnamefont {T.}~\bibnamefont
  {Grover}}, \bibinfo {author} {\bibfnamefont {D.~N.}\ \bibnamefont {Sheng}}, \
  and\ \bibinfo {author} {\bibfnamefont {A.}~\bibnamefont {Vishwanath}},\
  }\href {\doibase 10.1126/science.1248253} {\bibfield  {journal} {\bibinfo
  {journal} {Science}\ }\textbf {\bibinfo {volume} {344}},\ \bibinfo {pages}
  {280} (\bibinfo {year} {2014})},\ \Eprint {http://arxiv.org/abs/1301.7449}
  {arXiv:1301.7449 [cond-mat.str-el]} \BibitemShut {NoStop}%
\bibitem [{\citenamefont {Jian}\ \emph {et~al.}(2015)\citenamefont {Jian},
  \citenamefont {Jiang},\ and\ \citenamefont {Yao}}]{jian2015}%
  \BibitemOpen
  \bibfield  {author} {\bibinfo {author} {\bibfnamefont {S.-K.}\ \bibnamefont
  {Jian}}, \bibinfo {author} {\bibfnamefont {Y.-F.}\ \bibnamefont {Jiang}}, \
  and\ \bibinfo {author} {\bibfnamefont {H.}~\bibnamefont {Yao}},\ }\href
  {\doibase 10.1103/PhysRevLett.114.237001} {\bibfield  {journal} {\bibinfo
  {journal} {Phys. Rev. Lett.}\ }\textbf {\bibinfo {volume} {114}},\ \bibinfo
  {pages} {237001} (\bibinfo {year} {2015})}\BibitemShut {NoStop}%
\bibitem [{\citenamefont {Fei}\ \emph {et~al.}(2016)\citenamefont {Fei},
  \citenamefont {Giombi}, \citenamefont {Klebanov},\ and\ \citenamefont
  {Tarnopolsky}}]{Fei:2016sgs}%
  \BibitemOpen
  \bibfield  {author} {\bibinfo {author} {\bibfnamefont {L.}~\bibnamefont
  {Fei}}, \bibinfo {author} {\bibfnamefont {S.}~\bibnamefont {Giombi}},
  \bibinfo {author} {\bibfnamefont {I.~R.}\ \bibnamefont {Klebanov}}, \ and\
  \bibinfo {author} {\bibfnamefont {G.}~\bibnamefont {Tarnopolsky}},\ }\href
  {\doibase 10.1093/ptep/ptw120} {\bibfield  {journal} {\bibinfo  {journal}
  {PTEP}\ }\textbf {\bibinfo {volume} {2016}},\ \bibinfo {pages} {12C105}
  (\bibinfo {year} {2016})},\ \Eprint {http://arxiv.org/abs/1607.05316}
  {arXiv:1607.05316 [hep-th]} \BibitemShut {NoStop}%
\bibitem [{\citenamefont {Li}\ \emph {et~al.}(2017)\citenamefont {Li},
  \citenamefont {Jiang},\ and\ \citenamefont {Yao}}]{li2017edge}%
  \BibitemOpen
  \bibfield  {author} {\bibinfo {author} {\bibfnamefont {Z.-X.}\ \bibnamefont
  {Li}}, \bibinfo {author} {\bibfnamefont {Y.-F.}\ \bibnamefont {Jiang}}, \
  and\ \bibinfo {author} {\bibfnamefont {H.}~\bibnamefont {Yao}},\ }\href
  {\doibase 10.1103/PhysRevLett.119.107202} {\bibfield  {journal} {\bibinfo
  {journal} {Phys. Rev. Lett.}\ }\textbf {\bibinfo {volume} {119}},\ \bibinfo
  {pages} {107202} (\bibinfo {year} {2017})}\BibitemShut {NoStop}%
\bibitem [{\citenamefont {Deng}\ \emph {et~al.}(2005)\citenamefont {Deng},
  \citenamefont {Bl\"ote},\ and\ \citenamefont
  {Nightingale}}]{deng2005surface}%
  \BibitemOpen
  \bibfield  {author} {\bibinfo {author} {\bibfnamefont {Y.}~\bibnamefont
  {Deng}}, \bibinfo {author} {\bibfnamefont {H.~W.~J.}\ \bibnamefont
  {Bl\"ote}}, \ and\ \bibinfo {author} {\bibfnamefont {M.~P.}\ \bibnamefont
  {Nightingale}},\ }\href {\doibase 10.1103/PhysRevE.72.016128} {\bibfield
  {journal} {\bibinfo  {journal} {Phys. Rev. E}\ }\textbf {\bibinfo {volume}
  {72}},\ \bibinfo {pages} {016128} (\bibinfo {year} {2005})}\BibitemShut
  {NoStop}%
\bibitem [{Note4()}]{Note4}%
  \BibitemOpen
  \bibinfo {note} {Four fermion interaction, such as $\DOTSI \intop \ilimits@
  d^2 x (\psi ^\protect \dag \partial \psi )^2$, is irrelevant for the Majorana
  fermion}\BibitemShut {NoStop}%
\bibitem [{\citenamefont {{Fradkin}}(2013)}]{2013Fradkin}%
  \BibitemOpen
  \bibfield  {author} {\bibinfo {author} {\bibfnamefont {E.}~\bibnamefont
  {{Fradkin}}},\ }\href@noop {} {\emph {\bibinfo {title} {{Field Theories of
  Condensed Matter Physics}}}}\ (\bibinfo {year} {2013})\BibitemShut {NoStop}%
\bibitem [{\citenamefont {Wu}\ \emph {et~al.}(2006)\citenamefont {Wu},
  \citenamefont {Bernevig},\ and\ \citenamefont {Zhang}}]{wu2006}%
  \BibitemOpen
  \bibfield  {author} {\bibinfo {author} {\bibfnamefont {C.}~\bibnamefont
  {Wu}}, \bibinfo {author} {\bibfnamefont {B.~A.}\ \bibnamefont {Bernevig}}, \
  and\ \bibinfo {author} {\bibfnamefont {S.-C.}\ \bibnamefont {Zhang}},\ }\href
  {\doibase 10.1103/PhysRevLett.96.106401} {\bibfield  {journal} {\bibinfo
  {journal} {Phys. Rev. Lett.}\ }\textbf {\bibinfo {volume} {96}},\ \bibinfo
  {pages} {106401} (\bibinfo {year} {2006})}\BibitemShut {NoStop}%
\bibitem [{\citenamefont {Li}\ and\ \citenamefont {Yao}(2017)}]{li2017edgeb}%
  \BibitemOpen
  \bibfield  {author} {\bibinfo {author} {\bibfnamefont {Z.-X.}\ \bibnamefont
  {Li}}\ and\ \bibinfo {author} {\bibfnamefont {H.}~\bibnamefont {Yao}},\
  }\href {\doibase 10.1103/PhysRevB.96.241101} {\bibfield  {journal} {\bibinfo
  {journal} {Phys. Rev. B}\ }\textbf {\bibinfo {volume} {96}},\ \bibinfo
  {pages} {241101} (\bibinfo {year} {2017})}\BibitemShut {NoStop}%
\bibitem [{\citenamefont {Amit}\ \emph {et~al.}(1980)\citenamefont {Amit},
  \citenamefont {Goldschmidt},\ and\ \citenamefont {Grinstein}}]{Amit:1979ab}%
  \BibitemOpen
  \bibfield  {author} {\bibinfo {author} {\bibfnamefont {D.~J.}\ \bibnamefont
  {Amit}}, \bibinfo {author} {\bibfnamefont {Y.~Y.}\ \bibnamefont
  {Goldschmidt}}, \ and\ \bibinfo {author} {\bibfnamefont {G.}~\bibnamefont
  {Grinstein}},\ }\href {\doibase 10.1088/0305-4470/13/2/024} {\bibfield
  {journal} {\bibinfo  {journal} {J. Phys. A}\ }\textbf {\bibinfo {volume}
  {13}},\ \bibinfo {pages} {585} (\bibinfo {year} {1980})}\BibitemShut
  {NoStop}%
\bibitem [{\citenamefont {Raviv-Moshe}\ and\ \citenamefont
  {Zhong}(2023)}]{raviv-moshe2023phases}%
  \BibitemOpen
  \bibfield  {author} {\bibinfo {author} {\bibfnamefont {A.}~\bibnamefont
  {Raviv-Moshe}}\ and\ \bibinfo {author} {\bibfnamefont {S.}~\bibnamefont
  {Zhong}},\ }\href {\doibase 10.1007/JHEP08(2023)143} {\bibfield  {journal}
  {\bibinfo  {journal} {JHEP}\ }\textbf {\bibinfo {volume} {08}},\ \bibinfo
  {pages} {143} (\bibinfo {year} {2023})},\ \Eprint
  {http://arxiv.org/abs/2305.11370} {arXiv:2305.11370 [hep-th]} \BibitemShut
  {NoStop}%
\bibitem [{\citenamefont {Binnig}\ and\ \citenamefont
  {Rohrer}(1983)}]{BINNIG1983236}%
  \BibitemOpen
  \bibfield  {author} {\bibinfo {author} {\bibfnamefont {G.}~\bibnamefont
  {Binnig}}\ and\ \bibinfo {author} {\bibfnamefont {H.}~\bibnamefont
  {Rohrer}},\ }\href {\doibase https://doi.org/10.1016/0039-6028(83)90716-1}
  {\bibfield  {journal} {\bibinfo  {journal} {Surface Science}\ }\textbf
  {\bibinfo {volume} {126}},\ \bibinfo {pages} {236} (\bibinfo {year}
  {1983})}\BibitemShut {NoStop}%
\bibitem [{\citenamefont {Binnig}\ and\ \citenamefont
  {Rohrer}(1987)}]{Binnig1987}%
  \BibitemOpen
  \bibfield  {author} {\bibinfo {author} {\bibfnamefont {G.}~\bibnamefont
  {Binnig}}\ and\ \bibinfo {author} {\bibfnamefont {H.}~\bibnamefont
  {Rohrer}},\ }\href {\doibase 10.1103/RevModPhys.59.615} {\bibfield  {journal}
  {\bibinfo  {journal} {Rev. Mod. Phys.}\ }\textbf {\bibinfo {volume} {59}},\
  \bibinfo {pages} {615} (\bibinfo {year} {1987})}\BibitemShut {NoStop}%
\bibitem [{\citenamefont {Eggert}(2000)}]{eggert2000}%
  \BibitemOpen
  \bibfield  {author} {\bibinfo {author} {\bibfnamefont {S.}~\bibnamefont
  {Eggert}},\ }\href {\doibase 10.1103/PhysRevLett.84.4413} {\bibfield
  {journal} {\bibinfo  {journal} {Phys. Rev. Lett.}\ }\textbf {\bibinfo
  {volume} {84}},\ \bibinfo {pages} {4413} (\bibinfo {year}
  {2000})}\BibitemShut {NoStop}%
\bibitem [{\citenamefont {Kane}\ and\ \citenamefont {Fisher}(1992)}]{1992kane}%
  \BibitemOpen
  \bibfield  {author} {\bibinfo {author} {\bibfnamefont {C.~L.}\ \bibnamefont
  {Kane}}\ and\ \bibinfo {author} {\bibfnamefont {M.~P.~A.}\ \bibnamefont
  {Fisher}},\ }\href {\doibase 10.1103/PhysRevLett.68.1220} {\bibfield
  {journal} {\bibinfo  {journal} {Phys. Rev. Lett.}\ }\textbf {\bibinfo
  {volume} {68}},\ \bibinfo {pages} {1220} (\bibinfo {year}
  {1992})}\BibitemShut {NoStop}%
\bibitem [{\citenamefont {Li}\ \emph {et~al.}(2015)\citenamefont {Li},
  \citenamefont {Wang}, \citenamefont {Fu}, \citenamefont {Du}, \citenamefont
  {Schreiber}, \citenamefont {Mu}, \citenamefont {Liu}, \citenamefont
  {Sullivan}, \citenamefont {Cs\'athy}, \citenamefont {Lin},\ and\
  \citenamefont {Du}}]{li2015}%
  \BibitemOpen
  \bibfield  {author} {\bibinfo {author} {\bibfnamefont {T.}~\bibnamefont
  {Li}}, \bibinfo {author} {\bibfnamefont {P.}~\bibnamefont {Wang}}, \bibinfo
  {author} {\bibfnamefont {H.}~\bibnamefont {Fu}}, \bibinfo {author}
  {\bibfnamefont {L.}~\bibnamefont {Du}}, \bibinfo {author} {\bibfnamefont
  {K.~A.}\ \bibnamefont {Schreiber}}, \bibinfo {author} {\bibfnamefont
  {X.}~\bibnamefont {Mu}}, \bibinfo {author} {\bibfnamefont {X.}~\bibnamefont
  {Liu}}, \bibinfo {author} {\bibfnamefont {G.}~\bibnamefont {Sullivan}},
  \bibinfo {author} {\bibfnamefont {G.~A.}\ \bibnamefont {Cs\'athy}}, \bibinfo
  {author} {\bibfnamefont {X.}~\bibnamefont {Lin}}, \ and\ \bibinfo {author}
  {\bibfnamefont {R.-R.}\ \bibnamefont {Du}},\ }\href {\doibase
  10.1103/PhysRevLett.115.136804} {\bibfield  {journal} {\bibinfo  {journal}
  {Phys. Rev. Lett.}\ }\textbf {\bibinfo {volume} {115}},\ \bibinfo {pages}
  {136804} (\bibinfo {year} {2015})}\BibitemShut {NoStop}%
\bibitem [{\citenamefont {Maslov}\ and\ \citenamefont
  {Stone}(1995)}]{Landauer1995}%
  \BibitemOpen
  \bibfield  {author} {\bibinfo {author} {\bibfnamefont {D.~L.}\ \bibnamefont
  {Maslov}}\ and\ \bibinfo {author} {\bibfnamefont {M.}~\bibnamefont {Stone}},\
  }\href {\doibase 10.1103/PhysRevB.52.R5539} {\bibfield  {journal} {\bibinfo
  {journal} {Phys. Rev. B}\ }\textbf {\bibinfo {volume} {52}},\ \bibinfo
  {pages} {R5539} (\bibinfo {year} {1995})}\BibitemShut {NoStop}%
\bibitem [{\citenamefont {Maslov}(1995)}]{maslov1995}%
  \BibitemOpen
  \bibfield  {author} {\bibinfo {author} {\bibfnamefont {D.~L.}\ \bibnamefont
  {Maslov}},\ }\href {\doibase 10.1103/PhysRevB.52.R14368} {\bibfield
  {journal} {\bibinfo  {journal} {Phys. Rev. B}\ }\textbf {\bibinfo {volume}
  {52}},\ \bibinfo {pages} {R14368} (\bibinfo {year} {1995})}\BibitemShut
  {NoStop}%
\bibitem [{\citenamefont {Lezmy}\ \emph {et~al.}(2012)\citenamefont {Lezmy},
  \citenamefont {Oreg},\ and\ \citenamefont {Berkooz}}]{lezmy2012}%
  \BibitemOpen
  \bibfield  {author} {\bibinfo {author} {\bibfnamefont {N.}~\bibnamefont
  {Lezmy}}, \bibinfo {author} {\bibfnamefont {Y.}~\bibnamefont {Oreg}}, \ and\
  \bibinfo {author} {\bibfnamefont {M.}~\bibnamefont {Berkooz}},\ }\href
  {\doibase 10.1103/PhysRevB.85.235304} {\bibfield  {journal} {\bibinfo
  {journal} {Phys. Rev. B}\ }\textbf {\bibinfo {volume} {85}},\ \bibinfo
  {pages} {235304} (\bibinfo {year} {2012})}\BibitemShut {NoStop}%
\bibitem [{\citenamefont {Dzero}\ \emph {et~al.}(2010)\citenamefont {Dzero},
  \citenamefont {Sun}, \citenamefont {Galitski},\ and\ \citenamefont
  {Coleman}}]{dzero2010topological}%
  \BibitemOpen
  \bibfield  {author} {\bibinfo {author} {\bibfnamefont {M.}~\bibnamefont
  {Dzero}}, \bibinfo {author} {\bibfnamefont {K.}~\bibnamefont {Sun}}, \bibinfo
  {author} {\bibfnamefont {V.}~\bibnamefont {Galitski}}, \ and\ \bibinfo
  {author} {\bibfnamefont {P.}~\bibnamefont {Coleman}},\ }\href {\doibase
  10.1103/PhysRevLett.104.106408} {\bibfield  {journal} {\bibinfo  {journal}
  {Phys. Rev. Lett.}\ }\textbf {\bibinfo {volume} {104}},\ \bibinfo {pages}
  {106408} (\bibinfo {year} {2010})}\BibitemShut {NoStop}%
\bibitem [{\citenamefont {Alexandrov}\ \emph {et~al.}(2013)\citenamefont
  {Alexandrov}, \citenamefont {Dzero},\ and\ \citenamefont
  {Coleman}}]{alexandrov2013cubic}%
  \BibitemOpen
  \bibfield  {author} {\bibinfo {author} {\bibfnamefont {V.}~\bibnamefont
  {Alexandrov}}, \bibinfo {author} {\bibfnamefont {M.}~\bibnamefont {Dzero}}, \
  and\ \bibinfo {author} {\bibfnamefont {P.}~\bibnamefont {Coleman}},\ }\href
  {\doibase 10.1103/PhysRevLett.111.226403} {\bibfield  {journal} {\bibinfo
  {journal} {Phys. Rev. Lett.}\ }\textbf {\bibinfo {volume} {111}},\ \bibinfo
  {pages} {226403} (\bibinfo {year} {2013})}\BibitemShut {NoStop}%
\bibitem [{\citenamefont {Dzero}\ \emph {et~al.}(2016)\citenamefont {Dzero},
  \citenamefont {Xia}, \citenamefont {Galitski},\ and\ \citenamefont
  {Coleman}}]{dzero2016topological}%
  \BibitemOpen
  \bibfield  {author} {\bibinfo {author} {\bibfnamefont {M.}~\bibnamefont
  {Dzero}}, \bibinfo {author} {\bibfnamefont {J.}~\bibnamefont {Xia}}, \bibinfo
  {author} {\bibfnamefont {V.}~\bibnamefont {Galitski}}, \ and\ \bibinfo
  {author} {\bibfnamefont {P.}~\bibnamefont {Coleman}},\ }\href {\doibase
  https://doi.org/10.1146/annurev-conmatphys-031214-014749} {\bibfield
  {journal} {\bibinfo  {journal} {Annual Review of Condensed Matter Physics}\
  }\textbf {\bibinfo {volume} {7}},\ \bibinfo {pages} {249} (\bibinfo {year}
  {2016})}\BibitemShut {NoStop}%
\bibitem [{\citenamefont {Otrokov}\ \emph {et~al.}(2019)\citenamefont
  {Otrokov}, \citenamefont {Klimovskikh}, \citenamefont {Bentmann},
  \citenamefont {Estyunin}, \citenamefont {Zeugner}, \citenamefont {Aliev},
  \citenamefont {Ga{\ss}}, \citenamefont {Wolter}, \citenamefont {Koroleva},
  \citenamefont {Shikin} \emph {et~al.}}]{otrokov2019prediction}%
  \BibitemOpen
  \bibfield  {author} {\bibinfo {author} {\bibfnamefont {M.~M.}\ \bibnamefont
  {Otrokov}}, \bibinfo {author} {\bibfnamefont {I.~I.}\ \bibnamefont
  {Klimovskikh}}, \bibinfo {author} {\bibfnamefont {H.}~\bibnamefont
  {Bentmann}}, \bibinfo {author} {\bibfnamefont {D.}~\bibnamefont {Estyunin}},
  \bibinfo {author} {\bibfnamefont {A.}~\bibnamefont {Zeugner}}, \bibinfo
  {author} {\bibfnamefont {Z.~S.}\ \bibnamefont {Aliev}}, \bibinfo {author}
  {\bibfnamefont {S.}~\bibnamefont {Ga{\ss}}}, \bibinfo {author} {\bibfnamefont
  {A.}~\bibnamefont {Wolter}}, \bibinfo {author} {\bibfnamefont
  {A.}~\bibnamefont {Koroleva}}, \bibinfo {author} {\bibfnamefont {A.~M.}\
  \bibnamefont {Shikin}},  \emph {et~al.},\ }\href@noop {} {\bibfield
  {journal} {\bibinfo  {journal} {Nature}\ }\textbf {\bibinfo {volume} {576}},\
  \bibinfo {pages} {416} (\bibinfo {year} {2019})}\BibitemShut {NoStop}%
\bibitem [{\citenamefont {Bernevig}\ \emph {et~al.}(2022)\citenamefont
  {Bernevig}, \citenamefont {Felser},\ and\ \citenamefont
  {Beidenkopf}}]{bernevig2022progress}%
  \BibitemOpen
  \bibfield  {author} {\bibinfo {author} {\bibfnamefont {B.~A.}\ \bibnamefont
  {Bernevig}}, \bibinfo {author} {\bibfnamefont {C.}~\bibnamefont {Felser}}, \
  and\ \bibinfo {author} {\bibfnamefont {H.}~\bibnamefont {Beidenkopf}},\
  }\href@noop {} {\bibfield  {journal} {\bibinfo  {journal} {Nature}\ }\textbf
  {\bibinfo {volume} {603}},\ \bibinfo {pages} {41} (\bibinfo {year}
  {2022})}\BibitemShut {NoStop}%
\bibitem [{\citenamefont {Benalcazar}\ \emph {et~al.}(2017)\citenamefont
  {Benalcazar}, \citenamefont {Bernevig},\ and\ \citenamefont
  {Hughes}}]{benalcazar2017quantized}%
  \BibitemOpen
  \bibfield  {author} {\bibinfo {author} {\bibfnamefont {W.~A.}\ \bibnamefont
  {Benalcazar}}, \bibinfo {author} {\bibfnamefont {B.~A.}\ \bibnamefont
  {Bernevig}}, \ and\ \bibinfo {author} {\bibfnamefont {T.~L.}\ \bibnamefont
  {Hughes}},\ }\href {\doibase 10.1126/science.aah6442} {\bibfield  {journal}
  {\bibinfo  {journal} {Science}\ }\textbf {\bibinfo {volume} {357}},\ \bibinfo
  {pages} {61} (\bibinfo {year} {2017})},\ \Eprint
  {http://arxiv.org/abs/https://www.science.org/doi/pdf/10.1126/science.aah6442}
  {https://www.science.org/doi/pdf/10.1126/science.aah6442} \BibitemShut
  {NoStop}%
\bibitem [{\citenamefont {Schindler}\ \emph {et~al.}(2018)\citenamefont
  {Schindler}, \citenamefont {Cook}, \citenamefont {Vergniory}, \citenamefont
  {Wang}, \citenamefont {Parkin}, \citenamefont {Bernevig},\ and\ \citenamefont
  {Neupert}}]{Schindler_2018}%
  \BibitemOpen
  \bibfield  {author} {\bibinfo {author} {\bibfnamefont {F.}~\bibnamefont
  {Schindler}}, \bibinfo {author} {\bibfnamefont {A.~M.}\ \bibnamefont {Cook}},
  \bibinfo {author} {\bibfnamefont {M.~G.}\ \bibnamefont {Vergniory}}, \bibinfo
  {author} {\bibfnamefont {Z.}~\bibnamefont {Wang}}, \bibinfo {author}
  {\bibfnamefont {S.~S.~P.}\ \bibnamefont {Parkin}}, \bibinfo {author}
  {\bibfnamefont {B.~A.}\ \bibnamefont {Bernevig}}, \ and\ \bibinfo {author}
  {\bibfnamefont {T.}~\bibnamefont {Neupert}},\ }\href {\doibase
  10.1126/sciadv.aat0346} {\bibfield  {journal} {\bibinfo  {journal} {Science
  Advances}\ }\textbf {\bibinfo {volume} {4}} (\bibinfo {year} {2018}),\
  10.1126/sciadv.aat0346}\BibitemShut {NoStop}%
\bibitem [{\citenamefont {Harribey}\ \emph {et~al.}(2023)\citenamefont
  {Harribey}, \citenamefont {Klebanov},\ and\ \citenamefont
  {Sun}}]{Harribey:2023xyv}%
  \BibitemOpen
  \bibfield  {author} {\bibinfo {author} {\bibfnamefont {S.}~\bibnamefont
  {Harribey}}, \bibinfo {author} {\bibfnamefont {I.~R.}\ \bibnamefont
  {Klebanov}}, \ and\ \bibinfo {author} {\bibfnamefont {Z.}~\bibnamefont
  {Sun}},\ }\href {\doibase 10.1007/JHEP10(2023)017} {\bibfield  {journal}
  {\bibinfo  {journal} {JHEP}\ }\textbf {\bibinfo {volume} {10}},\ \bibinfo
  {pages} {017} (\bibinfo {year} {2023})},\ \Eprint
  {http://arxiv.org/abs/2307.00072} {arXiv:2307.00072 [hep-th]} \BibitemShut
  {NoStop}%
\bibitem [{\citenamefont {Sun}\ \emph {et~al.}(2023)\citenamefont {Sun},
  \citenamefont {Hu}, \citenamefont {Deng},\ and\ \citenamefont
  {Lv}}]{Sun:2023vwy}%
  \BibitemOpen
  \bibfield  {author} {\bibinfo {author} {\bibfnamefont {Y.}~\bibnamefont
  {Sun}}, \bibinfo {author} {\bibfnamefont {M.}~\bibnamefont {Hu}}, \bibinfo
  {author} {\bibfnamefont {Y.}~\bibnamefont {Deng}}, \ and\ \bibinfo {author}
  {\bibfnamefont {J.-P.}\ \bibnamefont {Lv}},\ }\href {\doibase
  10.1103/PhysRevLett.131.207101} {\bibfield  {journal} {\bibinfo  {journal}
  {Phys. Rev. Lett.}\ }\textbf {\bibinfo {volume} {131}},\ \bibinfo {pages}
  {207101} (\bibinfo {year} {2023})},\ \Eprint
  {http://arxiv.org/abs/2301.11720} {arXiv:2301.11720 [cond-mat.stat-mech]}
  \BibitemShut {NoStop}%
\bibitem [{Note5()}]{Note5}%
  \BibitemOpen
  \bibinfo {note} {A discussion of the boundary Gross-Neveu-Yukawa class in 3+1
  dimensions is provided in the Supplemental Material, where we obtain a new
  logarithmic correction to the boundary fermion correlation
  function.}\BibitemShut {Stop}%
\bibitem [{\citenamefont {Di~Pietro}\ \emph {et~al.}(2021)\citenamefont
  {Di~Pietro}, \citenamefont {Lauria},\ and\ \citenamefont
  {Niro}}]{dipietro20203d}%
  \BibitemOpen
  \bibfield  {author} {\bibinfo {author} {\bibfnamefont {L.}~\bibnamefont
  {Di~Pietro}}, \bibinfo {author} {\bibfnamefont {E.}~\bibnamefont {Lauria}}, \
  and\ \bibinfo {author} {\bibfnamefont {P.}~\bibnamefont {Niro}},\ }\href
  {\doibase 10.21468/SciPostPhys.11.3.050} {\bibfield  {journal} {\bibinfo
  {journal} {SciPost Phys.}\ }\textbf {\bibinfo {volume} {11}},\ \bibinfo
  {pages} {050} (\bibinfo {year} {2021})},\ \Eprint
  {http://arxiv.org/abs/2012.07733} {arXiv:2012.07733 [hep-th]} \BibitemShut
  {NoStop}%
\bibitem [{\citenamefont {Di~Pietro}\ \emph {et~al.}(2024)\citenamefont
  {Di~Pietro}, \citenamefont {Lauria},\ and\ \citenamefont
  {Niro}}]{dipietro2023conformal}%
  \BibitemOpen
  \bibfield  {author} {\bibinfo {author} {\bibfnamefont {L.}~\bibnamefont
  {Di~Pietro}}, \bibinfo {author} {\bibfnamefont {E.}~\bibnamefont {Lauria}}, \
  and\ \bibinfo {author} {\bibfnamefont {P.}~\bibnamefont {Niro}},\ }\href
  {\doibase 10.21468/SciPostPhys.16.4.090} {\bibfield  {journal} {\bibinfo
  {journal} {SciPost Phys.}\ }\textbf {\bibinfo {volume} {16}},\ \bibinfo
  {pages} {090} (\bibinfo {year} {2024})},\ \Eprint
  {http://arxiv.org/abs/2312.11633} {arXiv:2312.11633 [hep-th]} \BibitemShut
  {NoStop}%
\bibitem [{\citenamefont {Behan}\ \emph {et~al.}(2022)\citenamefont {Behan},
  \citenamefont {Di~Pietro}, \citenamefont {Lauria},\ and\ \citenamefont {van
  Rees}}]{behan2021bootstrapping}%
  \BibitemOpen
  \bibfield  {author} {\bibinfo {author} {\bibfnamefont {C.}~\bibnamefont
  {Behan}}, \bibinfo {author} {\bibfnamefont {L.}~\bibnamefont {Di~Pietro}},
  \bibinfo {author} {\bibfnamefont {E.}~\bibnamefont {Lauria}}, \ and\ \bibinfo
  {author} {\bibfnamefont {B.~C.}\ \bibnamefont {van Rees}},\ }\href {\doibase
  10.1007/JHEP03(2022)146} {\bibfield  {journal} {\bibinfo  {journal} {JHEP}\
  }\textbf {\bibinfo {volume} {03}},\ \bibinfo {pages} {146} (\bibinfo {year}
  {2022})},\ \Eprint {http://arxiv.org/abs/2111.04747} {arXiv:2111.04747
  [hep-th]} \BibitemShut {NoStop}%
\bibitem [{\citenamefont {Giombi}\ \emph {et~al.}(2016)\citenamefont {Giombi},
  \citenamefont {Klebanov},\ and\ \citenamefont
  {Tarnopolsky}}]{Giombi:2015haa}%
  \BibitemOpen
  \bibfield  {author} {\bibinfo {author} {\bibfnamefont {S.}~\bibnamefont
  {Giombi}}, \bibinfo {author} {\bibfnamefont {I.~R.}\ \bibnamefont
  {Klebanov}}, \ and\ \bibinfo {author} {\bibfnamefont {G.}~\bibnamefont
  {Tarnopolsky}},\ }\href {\doibase 10.1088/1751-8113/49/13/135403} {\bibfield
  {journal} {\bibinfo  {journal} {J. Phys. A}\ }\textbf {\bibinfo {volume}
  {49}},\ \bibinfo {pages} {135403} (\bibinfo {year} {2016})},\ \Eprint
  {http://arxiv.org/abs/1508.06354} {arXiv:1508.06354 [hep-th]} \BibitemShut
  {NoStop}%
\bibitem [{\citenamefont {Fei}\ \emph {et~al.}(2015)\citenamefont {Fei},
  \citenamefont {Giombi}, \citenamefont {Klebanov},\ and\ \citenamefont
  {Tarnopolsky}}]{Fei:2015oha}%
  \BibitemOpen
  \bibfield  {author} {\bibinfo {author} {\bibfnamefont {L.}~\bibnamefont
  {Fei}}, \bibinfo {author} {\bibfnamefont {S.}~\bibnamefont {Giombi}},
  \bibinfo {author} {\bibfnamefont {I.~R.}\ \bibnamefont {Klebanov}}, \ and\
  \bibinfo {author} {\bibfnamefont {G.}~\bibnamefont {Tarnopolsky}},\ }\href
  {\doibase 10.1007/JHEP12(2015)155} {\bibfield  {journal} {\bibinfo  {journal}
  {JHEP}\ }\textbf {\bibinfo {volume} {12}},\ \bibinfo {pages} {155} (\bibinfo
  {year} {2015})},\ \Eprint {http://arxiv.org/abs/1507.01960} {arXiv:1507.01960
  [hep-th]} \BibitemShut {NoStop}%
\end{thebibliography}%

\setcounter{secnumdepth}{3}
\setcounter{equation}{0}
\setcounter{figure}{0}
\renewcommand{\theequation}{S\arabic{equation}}
\renewcommand{\thefigure}{S\arabic{figure}}
\renewcommand\figurename{Supplementary Figure}
\renewcommand\tablename{Supplementary Table}
\newcommand\Scite[1]{[S\citealp{#1}]}
\makeatletter \renewcommand\@biblabel[1]{[S#1]} \makeatother

\onecolumngrid

\begin{center}
{\bf\large Supplemental Material}
\end{center}

\section{Mixed Yukawa fixed point and emergent Lorentz symmetry}

As a warm-up, we first consider a free scalar theory in the bulk coupled to a Dirac fermion in a codimension-1 boundary.
Note that other conformal boundary conditions for free scalar fields have been studied in Refs.~\cite{dipietro20203d,dipietro2023conformal,behan2021bootstrapping}. 
Such a theory can describe, for instance, the tricritical point in a 3+1D topological insulator. 
The scalar field lives in a $d$-dimensional half-infinite Euclidean spacetime, $\mathcal M$, spanned by $x^0\equiv \tau, x^1,...,x^{d-2} \in (-\infty, \infty)$ and $x^{d-1} \equiv y \in (0, \infty)$. 
While the Dirac fermion lives on the boundary of this half-infinite spacetime, $\mathcal{\partial M}$, with $\tau, x^1,...,x^{d-2} \in (-\infty, \infty)$, and $y=0$.
Note that $\tau$ stands for the imaginary time. 
The Euclidean action is
\begin{equation}
    S = \int_{\mathcal{M}} d^d x \frac{1}{2} \left((\partial_\tau\phi)^2+v_B^2 (\nabla \phi)^2 \right)+\int_{\partial \mathcal{M}} d^{d-1} x ( \bar{\psi}  (\gamma^0 \partial_0 + v_F \gamma \cdot \partial ) \psi - g \phi \bar{\psi} \psi) \,,
\end{equation}
where $\phi$, $\psi$ denotes the scalar field and the surface Dirac fermion with velocity $v_B$ and $v_F$, respectively. 
$g$ denotes the Yukawa interaction strength,
$\gamma \cdot \partial = \sum_{i=1}^{d-2} \gamma^i \partial_i $ with
$\{ \gamma^\mu, \gamma^\nu\}= 2\delta^{\mu\nu}$, $\mu, \nu= 0,...,d-2$, and $\bar \psi = \psi^\dag \gamma^0$.

Since the theory preserves translation symmetry in $x^{0,...,d-2}$ directions, we can work in the momentum space for those coordinates (frequency space for $\tau$), while in the real space for the $y$ coordinate. 
Notice that the fermion is localized on the surface, its free propagator is
\begin{equation}
    G_\psi(p) = i\frac{\gamma^0 \omega+v_F \gamma \cdot \bf{p}}{\omega^2+ v_F^2 {\bf p}^2} \,,
\end{equation}
where $\omega$ denotes the frequency. $\gamma \cdot {\bf p}  = \sum_{i=1}^{d-2} \gamma^{i} p_i$ with $p_i$ being the momentum in the $x^i$ coordinate, and ${\bf p}^2 \equiv \sum_{i=1}^{d-2} p_i^2$.

For the bulk boson, $(\partial_\tau^2+\sum_iv_B^2\partial_i^2)G_\phi = 0$, 
in the presence of a boundary at $y=0$, the propagator can be derived via the image method:
\begin{equation}
    G_{\phi}(x-x^\prime,y,y^\prime) = b_0\left(\frac{1}{\left(\left(x-x^{\prime}\right)^2+\left(y-y^{\prime}\right)^2\right)^{\frac{d-2}{2}}}+\frac{\lambda}{\left(\left(x-x^{\prime}\right)^2+\left(y+y^{\prime}\right)^2\right)^{\frac{d-2}{2}}}\right) \,.
\end{equation}
$b_0 = \frac{\Gamma(\frac{d-2}2)}{4\pi^{d/2}}$ is a coefficient that is set to make the coefficient of the propagator in the momentum space unity. 
Here $x = (v_B \tau, {\bf x})$ is a shorthand notation for coordinates parallel to the boundary. 
$\lambda$ is given by the boundary condition:
$\lambda= -1$ for Dirichlet boundary condition $\phi(x, y=0) = 0$ and $\lambda= 1$ for Neumann boundary condition $\partial_y\phi(x, y=0) = 0$. 
Here we will focus on the Neumann boundary condition $\lambda = 1$. 
After Fourier transformation for coordinates parallel to the boundary, the free propagator becomes
\begin{equation}
    G_{\phi}(p,y,y') = \frac{e^{-\sqrt{\omega^2+v_B^2 {\bf p}^2}|y-y'|} + e^{-\sqrt{\omega^2+v_B^2 {\bf p}^2}(y+y')}}{2\sqrt{\omega^2+v_B^2 {\bf p}^2}} \,.
\end{equation}
Since we are considering the surface criticality, we can restrict to $y = 0$. 
In this case, we define the free propagator for boson at the surface as
\begin{equation}
    G_{\phi}(p) = G_{\phi}(p,0,0) = \frac{1}{\sqrt{\omega^2+v_B^2 {\bf p}^2}} \,.
\end{equation}

The scaling dimension of Yukawa coupling is given by $[g] = \frac12(d-4)$.
Hence, we will work at $d=4-\epsilon$, and in this case, the boson self-interaction at the boundary $\int_{\partial \mathcal M } d^{d-1}x \phi^4$ is irrelevant. 
To conduct the one-loop renormalization group, we will consider the one-loop diagram for the self-energy of the boson and fermions respectively, and the Yukawa vertex. We will first assume the Lorentz symmetry by $v_B=v_F$ and set $v_B=1$ without loss of generality to obtain a stable fixed point.
And then we will show that at this stable fixed point the assumption is valid by calculating the RG flow of $v_B$ and $v_F$. 

To simplify the notation, we will use the following shorthand notation 
\begin{equation}
    G_\psi(p) = i\frac{\slashed{p}}{p^2} \,, \quad G_\phi(p) = \frac1{|p|} \,,
\end{equation}
where $\slashed{p} = \gamma^0 \omega +  \gamma \cdot \bf{p}$, $p^2 = \omega^2+ {\bf p}^2$ and $|p|= \sqrt{p^2}$.

\begin{figure}
\centering\includegraphics[width = 1.0\textwidth]{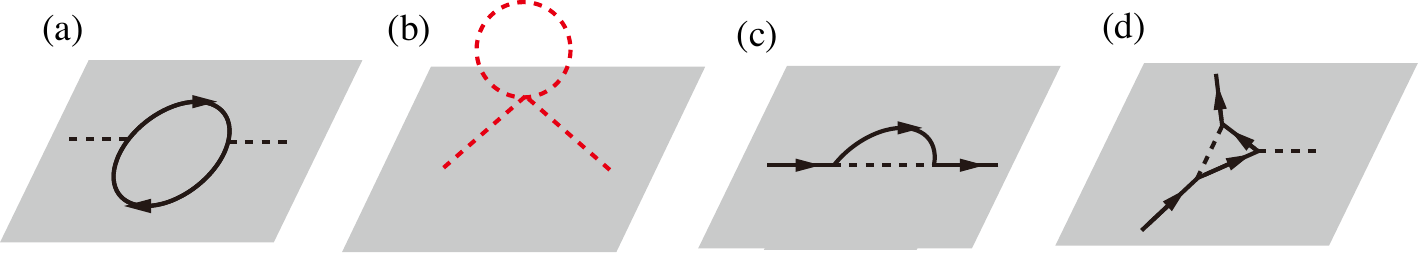}
    \caption{(a) Scalar one loop propagator
correction; (b) Fermion one loop propagator correction; (c) One loop vertex correction. The Black dashed line represents the projected boson. The Red dashed line represents the bulk boson. Balck solid arrow represents the fermion.}
    \label{fig:enter-label}
\end{figure}

The self-energy for the boson is
\begin{equation}
    \Pi_{\phi}(q) = (-1)g^2 \int \frac{d^{d-1} p}{(2 \pi)^{d-1}} \frac{\operatorname{tr}[i \slashed{p} i(\slashed{p}+\slashed{q})]}{p^2(p+q)^2} = g^2 \frac{2^{5-2 d} \pi^{2-\frac{d}{2}}}{\cos \left(\frac{\pi d}{2}\right) \Gamma\left(\frac{d}{2}-1\right)} q^{d-3} \,.
\end{equation}
Here we choose $\operatorname{tr}(\gamma^\mu\gamma^\nu) = 2\delta^{\mu\nu}$.
There is no singularity at $d=4-\epsilon$ for $\phi$ and thus $Z_\phi$ should be finite. 
The self-energy for the fermion is
\begin{equation}
    \begin{aligned}
    \Pi_\psi(q) & = g^2 \int \frac{d^{d-1} p}{(2 \pi)^{d-1}} \frac{(i \slashed{p})}{p^2|p-q|} \\
    & = i g^2 \frac{4^{2-d} \pi^{\frac{1-d}{2}} \Gamma\left(2-\frac{d}{2}\right) \Gamma(d-2)}{\Gamma\left(d-\frac{3}{2}\right)} \frac{\slashed q}{q^{4-d}} \,, 
\end{aligned}
\end{equation}
and in $d = 4-\epsilon$, the singular terms can be extracted,
\begin{equation}
    \Pi_\psi(q)=i \slashed{q} g^2\left[\frac{1}{6 \pi^2 \epsilon}+\frac{1}{36 \pi^2}\left(10-3 \gamma-3 \log \left(q^2 / \pi\right)\right)\right]\,.
\end{equation}
The one-loop vertex is given by:
\begin{equation}\label{lorentz_symmetry_self_psi}
\Gamma\left(k, q\right)= g^3 \int \frac{d^{d-1} p}{(2 \pi)^{d-1}} \frac{i(\slashed{p}+\slashed{k}) i(\slashed{p}+\slashed{q})}{(p+k)^2(p+q)^2|p|}\approx -g^3 \frac{1}{2 \pi^2 \epsilon}+\text { finite }.
\end{equation}
$Z$ factors can then be extracted via this one-loop calculation:
\begin{equation}
\begin{aligned}
& Z_\psi=1-g^2\frac{1}{6 \pi^2 \epsilon} \,, \\
& Z_\phi=1+g^2(\mathcal{O}(1)) \,, \\
& Z_g=1+g^2\frac{1}{2 \pi^2 \epsilon} \,.
\end{aligned}
\end{equation}
The bare coupling is associated with renormalized coupling via
\begin{equation}
g_{\text{bare}} Z_\phi^{1 / 2} Z_\psi=g \mu^{\epsilon / 2} Z_g\,.
\end{equation}

This gives the beta function
\begin{equation}
\beta=-\frac{\epsilon}{2} g+\frac{2}{3 \pi^2} g^3+\mathcal{O}\left(g^4\right)\,.
\label{rg:free}
\end{equation}
For $\epsilon > 0$, the Gaussian fixed point is unstable, and there is a stable nontrivial IR fixed point $g_*^2 = \frac{3 \pi^2}{4} \epsilon$.

Let us now include the renormalization of velocity, i.e. we consider $v_{F}\neq v_B$. 
Again, we calculate the contribution of the three one-loop diagrams. 
The one-loop self-energy for bosons is
\begin{equation}
\begin{aligned}
   &  \Pi_{\phi}(\nu,{\bf p}) = 2 g^2 \int \frac{d^{d-2}  q}{(2 \pi)^d} \int_{-\infty}^{\infty} \frac{d \omega}{2 \pi} \frac{\omega(\omega+\nu)+v_F^2 \textbf q_j (\textbf p_j+\textbf q_j)}{\left(\omega^2+v_F^2 \textbf q^2\right)\left[(\omega+\nu)^2+v_F^2(\textbf p+\textbf q)^2\right]}\\
    & =\frac{g^2}{2} \int_0^1 d x\int \frac{d^{d-2} q}{(2 \pi)^d}\left[\frac{v_F^2 \textbf q^2+x(1-x)\left(\nu^2+v_F^2 \textbf p^2\right)-x(1-x) \nu^2+v_F^2 \textbf q^2-x(1-x) v_F^2 \textbf p^2}{\left[v_F^2 \textbf q^2+x(1-x)\left(\nu^2+v_F^2 \textbf p^2\right)\right]^{3 / 2}}\right]\\
    & = \frac{2\Gamma(\frac{5}{2}-d)(d-2)g^2}{(4\pi)^{\frac{d-1}{2}}v_F^{d-2}}\int_0^1dx\frac{1}{\left[x(1-x)(\nu^2+v_F^2\textbf p^2)\right]^{\frac{3}{2}-d}}\\
    & =-\frac{g^2}{8 v_F^2}\sqrt{\nu^2+v_F^2\textbf p^2}+\mathcal{O}(\epsilon) \,.
    \end{aligned}
\end{equation}
The one-loop self-energy for fermions is
\begin{equation}
\begin{aligned}
    \Pi_{\psi}(\nu,\textbf q) & = ig^2\int \frac{dp^{d-2}}{(2\pi)^{d-2}}\int \frac{d\omega}{2\pi}\frac{\omega\gamma_0+v_F\textbf p_j\gamma^j}{(\omega^2+v_F^2\textbf p^2)\sqrt{(\omega+\nu)^2+v_B^2(\textbf p+\textbf q)^2}}\\
    & = \frac{ig^2}{2}\int \frac{dp^{d-2}}{(2\pi)^{d-2}}\int \frac{d\omega}{2\pi}\int_0^1 dx\frac{\omega\gamma_0+v_F\textbf p_j\gamma^j}{\left[(1-x)(\omega^2+v_F^2\textbf p^2)+x((\omega+\nu)^2+v_B^2(\textbf p+\textbf q)^2)\right]^\frac{3}{2}} \,.
\end{aligned}
\end{equation}
To proceed, let $\textbf p+\frac{x v_B^2}{(1-x)v_F^2+x v_B^2} \textbf q \rightarrow \textbf p$, $\omega + x \nu \rightarrow \omega$ and rearrange the integral, we have
\begin{equation}
\begin{aligned}
     &\frac{ig^2}{2}\int \frac{dp^{d-2}}{(2\pi)^{d-2}}\int \frac{d\omega}{2\pi}\int_0^1 \frac{dx}{\sqrt{x}}\frac{\omega\gamma_0+v_Fp_j\gamma_j}{\left[(1-x)(\omega^2+v_F^2p^2)+x((\omega+\nu)^2+v_B^2(p+q)^2)\right]^\frac{3}{2}}\\
     =&\frac{ig^2}{2}\int \frac{dp^{d-2}}{(2\pi)^{d-2}}\int \frac{d\omega}{2\pi}\int_0^1 \frac{dx}{\sqrt{x}}\frac{(\omega-x\nu)\gamma_0+v_F\dk{p_j-\frac{xv_B^2}{(1-x)v_F^2+xv_B^2}q_j}\gamma_j}{\left[\omega^2+x(1-x)\nu^2+((1-x)v_F^2+xv_B^2)p^2+\frac{x(1-x)v_B^2v_F^2}{((1-x)v_F^2+xv_B^2)}q^2)\right]^\frac{3}{2}}\\
     =& \frac{ig^2}{2\pi}\int \frac{dp^{d-2}}{(2\pi)^{d-2}}\int_0^1 \frac{dx}{\sqrt{x}}\frac{-x\nu\gamma_0+v_F\dk{p_j-\frac{xv_B^2}{(1-x)v_F^2+xv_B^2}q_j}\gamma_j}{\left[x(1-x)\nu^2+((1-x)v_F^2+xv_B^2)p^2+\frac{x(1-x)v_B^2v_F^2}{((1-x)v_F^2+xv_B^2)}q^2)\right]}\\
     =&ig^2\frac{2\Gamma\dk{2-\frac{d}{2}}}{(4\pi)^{\frac{d}{2}}}\int_0^1 \frac{dx}{\sqrt{x}}\frac{-x\nu\gamma_0+v_F\dk{-\frac{xv_B^2}{(1-x)v_F^2+xv_B^2}q_j}\gamma_j}{\left[\frac{x(1-x)\nu^2}{(1-x)v_F^2+xv_B^2}+\frac{x(1-x)v_B^2v_F^2}{((1-x)v_F^2+xv_B^2)^2}q^2\right]^{2-\frac{d}{2}}((1-x)v_F^2+xv_B^2)}\\
     =&-\frac{ig^2}{\epsilon}\left[\frac{\nu\gamma_0\dk{\sqrt{v_B^2-v_F^2}-v_F\arctan(\frac{\sqrt{v_B^2-v_F^2}}{v_F}}}{2\pi^2(v_B^2-v_F^2)^{\frac{3}{2}}}-\frac{q_j\gamma_j\dk{v_F-\frac{v_B^2\arctan\frac{\sqrt{v_B^2-v_F^2}}{v_F}}{\sqrt{v_B^2-v_F^2}}}}{4\pi^2v_F(v_B^2-v_F^2)}\right]+\mathcal{O}(1)\\
     =& -\frac{ig^2}{4\pi^2\epsilon}\left[\nu\gamma_0\dk{\frac{2}{(v_B^2-v_F^2)}-\frac{2v_F\arctan\frac{\sqrt{v_B^2-v_F^2}}{v_F}}{(v_B^2-v_F^2)^\frac{3}{2}}}+q_j\gamma_j\dk{-\frac{1}{(v_B^2-v_F^2)}+\frac{v_B^2\arctan\frac{\sqrt{v_B^2-v_F^2}}{v_F}}{v_F(v_B^2-v_F^2)^{\frac{3}{2}}}}\right]+\mathcal{O}(1) \,.
    \end{aligned}
\end{equation}
It is direct to verify that when $v_F=v_B = 1$, the result nicely coincides with Eq.~\eqref{lorentz_symmetry_self_psi}. 
Now, we can directly extract the $Z$ factors:
\begin{equation}
    \begin{aligned}
        Z_\phi & = 1+g^2\mathcal{O}(1) \,, \\
        Z_{v_B} & = Z_{\phi}^{-\frac{1}{2}}\dk{1+g^2\mathcal{O}(1)} \,,  \\
        Z_\psi & = 1-\frac{g^2}{4\pi^2\epsilon}\dk{\frac{2}{(v_B^2-v_F^2)}-\frac{2v_F\arctan\frac{\sqrt{v_B^2-v_F^2}}{v_F}}{(v_B^2-v_F^2)^\frac{3}{2}}}+\mathcal{O}(1) \,, \\
        Z_{v_F} & = Z_\psi^{-1}\dk{1-\frac{g^2}{4\pi^2\epsilon}\dk{\frac{1}{(v_B^2-v_F^2)}-\frac{v_B^2\arctan\frac{\sqrt{v_B^2-v_F^2}}{v_F}}{v_F(v_B^2-v_F^2)^{\frac{3}{2}}}}+\mathcal{O}(1)} \,.
    \end{aligned}
\end{equation}
$Z_\phi$ and $Z_{v_B}$ are both finite, which means that the boson velocity does not receive the renormalization from the boundary. 
With
\begin{equation}
v_{F,\text{bare}}= Z_{v_F}Z_{\psi}^{-1}v_F \,,
\end{equation}
the beta function for $v_F$ can be evaluated as
\begin{equation}
    \beta_{v_F} = -\frac{3g^2}{8\pi^2(v_B^2-v_F^2)}+\frac{g^2(2v_F+\frac{v_B^2}{v_F})\arctan\frac{\sqrt{v_B^2-v_F^2}}{v_F}}{8\pi^2(v_B^2-v_F^2)^{\frac{3}{2}}} \,.
\end{equation}
At the stable fixed point $g_*^2$, the fixed point value of $v_F$ can be directly solved to be $v_B$, which indicates an emergent Lorentz symmetry.
Therefore, it justifies our assumption.

Now setting $v_F = v_B = 1$, the boundary field $\psi$ acquires an anomalous dimension,
\begin{equation}
\hat{\eta}_\psi=\mu \frac{\partial}{\partial \mu} \log Z_\psi^{\frac{1}{2}} = \frac{g^2}{12\pi^2}  =  \frac{\epsilon}{16} \,,
\end{equation}
in the last step, we have substituted the fixed point value $g^2_\ast$.

\section{Boundary Gross-Neveu-Yukawa Universality class}

In this section, we generalize our previous calculation to include a self-interaction for the bulk scalar field.
The corresponding Euclidean action reads
\begin{equation}
    S = \int_{} d^{d}x \left(\frac{1}{2}(\partial\phi)^2+\frac{1}{4!}\lambda\phi^4\right) + \int_{} d^{d-1}x\dk {\bar{\psi}\gamma^\mu\partial_\mu\psi-g\phi\bar{\psi}\psi},
\end{equation}
where $\gamma^\mu \partial_\mu = \sum_{i=0}^{d-2} \gamma^i \partial_i$ and $\int d^dx \phi^4$ is the self-interaction for the bulk boson with strength given by $\lambda$. 
Notice that we again assume the Lorentz symmetry, and will justify this assumption later.

The scaling dimensions are $[\lambda] = 4-d,[g] = 2-\frac{d}{2}$. 
Hence, for $d = 4$, the boson self-interaction $\phi^4$ and the Yukawa coupling are both tree-level marginal, which allows us to perform a RG calculation at $d = 4 - \epsilon$. 
For the same reason, we neglect the surface $\int_{\partial \mathcal M} d^{d-1} x \phi^4$. 
According to the previous study, while the boundary theory will not renormalize from the bulk theory, the bulk theory can lead to nontrivial correction to the boundary theory. 
At one-loop level, no Feynman graphs involved both Yukawa vertex and $\phi^4$ vertex exist. 
However, the $\phi^4$ vertex will contribute to the self-energy of the boundary boson by the tadpole diagram which we denote $\Pi_{\phi}^{\phi^4}$ and result in a surface anomalous dimension,
\begin{equation}
\begin{aligned}
    \Pi^{\phi^4}_{\phi}(p) & = \frac{\lambda}{2}\int \frac{d^{d-1}k}{(2\pi)^{d-1}}\int_0^\infty dy\frac{e^{-|p|y}}{|p|}\frac{1+e^{-2|k|y}}{2|k|}\frac{e^{-|p|y}}{|p|}\\
    & = \frac{\lambda}{8}\int \frac{d^{d-1}k}{(2\pi)^{d-1}}\dk{\frac{1}{|p|^3|k|}+\frac{1}{(|k|+|p|)|k||p|^2}}\\
    & = \frac{\lambda}{8}\frac{2\pi^{\frac{d-1}{2}}}{\Gamma(\frac{d-1}{2})}\int\frac{dk}{(2\pi)^{d-1}}\dk{\frac{ |k|^{d-3}}{|p|^3}+\frac{|k|^{d-3}}{(|p|+|k|)|p|^2}} \,.
    \end{aligned}
\end{equation}
The superscript indicates the contribution from the $\phi^4$ vertex. 
According to Veltman's formula, the first term vanishes and the second term is
\begin{equation}
    \frac{\lambda}{4}\frac{\pi^{\frac{d-1}{2}}}{\Gamma(\frac{d-1}{2})(2\pi)^{d-1}}\frac{1}{p^{5-d}}\int_0^1duu^{2-d}(1-u)^{d-3} = \frac{\lambda}{4}\frac{\Gamma(3-d)\Gamma(d-2)\pi^{\frac{d-1}{2}}}{\Gamma(1)\Gamma(\frac{d-1}{2})(2\pi)^{d-1}}\frac{1}{p^{5-d}} \,,
\end{equation}
where we set $u = \frac{|p|}{|k|+|p|}$. 
Expanding as a series of $\epsilon$, we arrive at
\begin{equation}
    G_{\phi}^{\phi^4}(p) = -\frac{\lambda}{16\pi^2|p|}\frac{1}{\epsilon}\,.
\end{equation}
This modifies $Z_{\phi}$
\begin{equation}
    Z_{\phi} = 1+\frac{\lambda}{16\pi^2}\frac{1}{\epsilon} \,. 
\end{equation}
Hence, the modified $\beta$ function for Yukawa coupling $g$ and the $\beta$ function for $\lambda$, which gives the usual Wilson-Fisher fixed point, are respectively given by:
\begin{equation}
    \begin{split}
    & \beta_g = -\frac{\epsilon}{2}g+\frac{2}{3\pi^2}g^3-\frac{1}{32\pi^2}\lambda g \,, \\ 
    & \beta_{\lambda} = -\epsilon\lambda +\frac{3}{(4\pi)^2}\lambda^2 \,.
    \end{split}
\end{equation}
From the second equation, it is clear that the IR fixed point of $\lambda$ is given by the usual Wilson-Fisher point $\lambda_* = \frac{(4\pi)^2\epsilon}{3}$ by the bulk theory.
Back to the first equation, we find a new fixed point for $g$ at $g_*^2 = \pi^2\epsilon$. We call this new fixed point the boundary Gross-Neveu-Yukawa fixed point.
It is a stable fixed point from the Jacobian
\begin{equation}
    M_{i j}=\frac{\partial \beta_i}{\partial g_j}, \quad M=\left[\begin{array}{cc}-\epsilon+\frac{3}{8\pi^2}\lambda_* & 0 \\ \frac{1}{32\pi^2}g_* & -\frac{\epsilon}{2}-\frac{1}{32\pi^2}\lambda_*+\frac{2}{\pi^2}g_*^{2}\end{array}\right] \,,
\end{equation}
whose eigenvalues, i.e., $\frac{4}{3}\epsilon, \epsilon$, are all positive, meaning that this fixed point is stable. 
At this new fixed point, the fermion anomalous dimensions are
\begin{equation}
    \eta_{\psi} = \frac{g_*^2}{12\pi^2} = \frac{\epsilon}{12} \,.
\end{equation}

\subsection{Logrithmic boundary flow in 3+1 dimensions}

Our results of boundary Gross-Neveu-Yukawa universality class apply to the 3+1 dimensional systems with 2+1 dimensional boundaries. 
Setting $\epsilon=0$, the beta equations in 3+1 dimensions read 
\begin{eqnarray} \label{eq:bGNY}
    \beta_g & = & \frac{2}{3\pi^2}g^3 -\frac{1}{32\pi^2}\lambda g \,,\\
\label{eq:ising_bulk}       
    \beta_\lambda & = & \frac{3}{16\pi^2}\lambda^2 \,.
\end{eqnarray}
Apparently, in the IR, the coupling constants exhibit a logarithmic flow as shown below
\begin{eqnarray}
    \lambda \approx \frac{16\pi^2}{3 \log \Lambda/\mu}\,, \quad g^2 \approx \frac{\pi^2}{\log \Lambda/\mu}\, ,
\end{eqnarray}
in which $\Lambda$ is the UV cutoff and $\mu \ll \Lambda$ is the IR energy scale.
We keep the leading terms.
The logarithmic flow will lead to a logarithmic correction to the boundary fermion correlation functions.
Consider the Callan-Symanzik equation for the boundary fermion correlation function,
\begin{eqnarray}
    \left( \mu \frac{\partial }{\partial \mu} + \beta_g \frac{\partial}{\partial g} + \beta_\lambda \frac{\partial}{\partial \lambda} +  \eta_{\psi}   \right) G_{\psi}(r;\mu, g, \lambda) = 0,
\end{eqnarray}
where $G_{\psi}$ denotes the two-point function of the boundary fermions, and the fermion anomalous dimension is
\begin{eqnarray}
    \eta_\psi = \frac{g^2}{12\pi^2} \approx \frac1{12} \frac1{\log \Lambda/\mu}\, . 
\end{eqnarray}
We can solve the Callan-Symanzik equation to get the renormalized correlation at the IR,
\begin{eqnarray}
    G_\psi(r) = \frac1{r^2} \left( \log \Lambda r\right)^{\frac16},
\end{eqnarray}
where the logarithmic correction is universal to the boundary fermions. 

\section{Special BKT phase transition}

In this section, we consider the 3-dimensional Ising boundary criticality in the presence of boundary fermion. 
As explained in the main text, our strategy is to perturb the ordinary fixed point by coupling with a boundary fermion. 
After bosonization of the Yukawa coupling between the Ising field and the boundary fermion, 
\begin{equation}
    S_{\text{f}} = \int dxd\tau\left[\frac{1}{2K}\left(\frac{1}{v_f}(\partial_\tau\varphi)^2+v_f(\partial_x\varphi)^2\right)+g\phi\cos(\sqrt{4\pi}\varphi)\right]
\end{equation}
where $\phi$ denotes the Ising field with scaling dimension $\Delta_{\hat \phi}$, and $\varphi$ denotes the bosonization field. 
$K$ and $g$ are the Luttinger parameter and the Yukawa coupling, respectively, and the velocity of the fermion is $v_f$. 
Without loss of generality, we can rescale the bulk boson velocity to be $v_b = 1$, the velocity ratio $v_f = v_f/v_b$ then enters into the RG flow of velocity. 
The OPE of the vertex operator is now
\begin{equation}
    \begin{aligned}
  &:\phi V_1(x)::\phi V_{-1}(0):  = \\ 
  & \frac{C_{\phi}}{\left(x^{2}+v_{f}^{2} \tau^{2}\right)^{K }\left(x^{2}+ \tau^{2}\right)^{\Delta_{\phi}}}\left[1+i\sqrt{4\pi}\left(x \partial_{x} \varphi(0)+\tau \partial_{\tau} \varphi(0)\right)\right. \left.-\frac{4\pi}{2}\left(x^{2}\left(\partial_{x} \varphi\right)^{2}+\tau^{2}\left(\partial_{\tau} \varphi\right)^{2}+2 x \tau \partial_{x} \varphi \partial_{\tau} \varphi\right)\right]
    \end{aligned}
\end{equation}
where $\Delta_{\hat\phi}$ is the scaling dimension of the boundary boson $\phi$, and $b$ is the coefficient in the correlation function of the boundary boson: $G_{\phi}(x)=\frac{C_{\phi}}{x^{2\Delta_{\hat\phi}}}$. 
For simplicity, we choose to renormalize the two-point function of $\phi$, namely set $C_{\phi} = 1$. 
The consistent OPE coefficients can be derived from the standard conformal perturbation theory \cite{Giombi:2015haa,Fei:2015oha}. 
We need not worry about the issue of normalization, since the results can be related by properly rescaling $g$.
The renormalized action takes the form of 
\begin{equation}
\begin{split}
    \delta S&=\frac{g^2}{8} \int d x d \tau \int_{a^{2}<x^{\prime 2}+ \tau^{\prime 2}<a^{2}(1+2 d \ell)} d x^{\prime} d \tau^{\prime}\left[\left(x^{\prime 2}(\partial \varphi(x))^{2}+\tau^{\prime 2}\left(\partial_{\tau} \varphi(x)\right)^{2}\right) \frac{4\pi }{\left(x^{\prime 2}+v_{f}^{2} \tau^{\prime 2}\right)^{K}\left(x^{\prime 2}+ \tau^{\prime 2}\right)^{\Delta_{\phi}}}\right]\\
    &=dlg^2\int d x d \tau \left[2\pi v_fA(v_f)(\partial \varphi(x))^{2}+\frac{2\pi }{v_f}B(v_f)\left(\partial_{\tau} \varphi(x)\right)^{2}\right]\, ,
    \end{split}
\end{equation}
where the two functions $A(v_f), B(v_f)$ are defined as 
\begin{equation}
    A\left(v_{f} \right)=\frac{1}{4 v_{f}} \int_{0}^{2 \pi} d \theta \frac{\cos ^{2} \theta}{\left(\cos ^{2} \theta+v_{f}^{2} \sin ^{2} \theta\right)^{K} }\, , \qquad  B\left(v_{f} \right)=\frac{v_{f}}{4 } \int_{0}^{2 \pi} d \theta \frac{\sin ^{2} \theta}{\left(\cos ^{2} \theta+v_{f}^{2} \sin ^{2} \theta\right)^{K}}\, .
\end{equation}
So we have the following RG equations
\begin{equation}
    \begin{aligned}
        &\frac{dK}{dl}=-2\pi\left(A(v_f)+B(v_f)\right)  g^{2}K^2 \,,\\
        &\frac{dv_f}{dl}=2\pi\left(A(v_f)-B(v_f)\right)  g^{2}v_fK \,,\\
        &\frac{dg}{dl}=(2-\Delta_{\hat \phi}-K)g\,.
    \end{aligned}
\end{equation}

We can check the RG equations at $v_f=1$, and now $A(v_f=1)=B(v_f=1)=\frac{\pi}{4}$. If we consider the boundary of $2+1$d TIs, then $\Delta_{\hat{\phi}} = \Delta_{\text{ord}}$, and we have:
\begin{equation}
    \begin{aligned}
        &\frac{dK}{dl}=-  \pi^2g^{2}K^2 \,,\\
        &\frac{dg}{dl}=(2-\Delta_{\hat{\phi}}-K)g \,.
    \end{aligned}
\end{equation}
Considering the two-particle backscattering term $g_2\cos(\sqrt{16\pi}\varphi)$, the RG equation yields 
\begin{equation}
        \begin{aligned}
        &\frac{dK}{dl}=-  \pi^2g^{2}K^2-4\pi^2 g_2^2K^2\,,\\
        &\frac{dg}{dl}=(2-\Delta_{\hat{\phi}}-K)g\,,
        \\
        & \frac{dg_2}{dl} = (2-4K)g_2\,.
    \end{aligned}
\end{equation}
For the defect case, the coefficients of the projected boson propagator should be twice larger than the boundary case. 
And as we mentioned before, it can be simply incorporated by redefining $g$ without changing the physics.

\section{Higher order special BKT phase transition}
In this section, we evaluate the RG flow of the special BKT phase transition up to the third order \cite{Amit:1979ab}. 
We start with the action
\begin{equation}
    S=\int d^2 x\left[\frac{1}{2 }(\partial \varphi)^2+\frac{g}{K}\phi \cos (\sqrt{K}\varphi)\right]+ S_{\text{bulk}}\,,
\end{equation}
where $S_{\text{bulk}}$ is the action of $\phi$ in the bulk. 
Notice here, we have redefined the field and coupling constant for simplicity as compared to the action in the main text. 
We choose such a convention for the reason that the log divergence can be absorbed into two independent renormalization factors $Z_g, Z_{\varphi}$. 

\begin{equation}
    \begin{aligned}\label{Zfactor}
        g_\text{bare} = Z_g g \,, \quad 
        K_\text{bare} = Z_\varphi^{-1} K \,.
    \end{aligned}
\end{equation}
\begin{figure}[t]
    \centering
    \includegraphics[width = 0.65\textwidth]{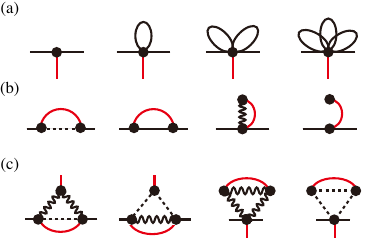}
    \caption{(a) Feynman diagram for the zero-order vertex correction $\Gamma^{(0)}$ ($g\phi\varphi^2$ as example and up to three loops). The red external line is the $\phi$ entry and black is the $\varphi$ entry. The black solid vertex represents the dressed vertex. The black solid internal line is the propagator of $\varphi$. (b) Feynman diagram for the second order self-energy $\Sigma^{(2)}_\varphi(p)$. The dashed line represents $\cosh G_\varphi-1$ and the wavy line represents $\sin G_\varphi-G_\varphi$. (c) \textit{Typical} Feynman diagrams that contribute to the log divergence in the third order of vertex correction $\Gamma^{(3)}_g$. Diagrams related by symmetry operations are neglected here. The complete expressions of the diagrams are listed in Eq.~\eqref{eq::gamma3}.}
    \label{fig:hofd}
\end{figure}

The propagator of $\sqrt{K}\varphi$ is
\begin{equation}
    G_\varphi(x, a)=K\left.\int \frac{d^2 p}{(2 \pi)^2} \frac{\mathrm{e}^{\mathrm{i} p y}}{p^2+m_0^2}\right|_{y^2=x^2+a^2}=\frac{K}{2 \pi} K_0\left[m_0\left(x^2+a^2\right)^{1 / 2}\right] \sim \frac{K}{4\pi}\ln\left(x^2+a^2\right)\, ,
\end{equation}
where $m_0$ is the IR cutoff and $a$ is the UV cutoff. In the case of a small IR cutoff, the propagator behaves logarithmically. $K_0(r)$ is the second-type modified Bessel function. Notice that the $\phi$ is not normalized by $\varphi$, we first trigger the flow of the bulk to infrared and then consider its effect on the surface. The $\phi$ propagator is chosen to be $\frac{1}{(x^2+a^2)^{2\Delta_{\phi}}}$. 
Our strategy is to expand the vertex $\frac{g}{K}\phi\cos(\sqrt{K}\varphi) = \sum_{n = 0}^{\infty} \frac{g}{K}\phi\frac{(-1)^n(\sqrt{K}\varphi)^{2n}}{(2n)!}$ and resum all the log-divergence up the higher order of $g$. We first focus on the tree-level of $g$, which originates from the vertex correction illustrated in Fig.~\ref{fig:hofd} (a). After resummation, the correction is
\begin{equation}
    g_\text{bare} = \left.a^{\Delta_{\hat{\phi}}-2}\exp\left(-\frac{1}{2}G_{\varphi}(x)\right)g\right|_{x = 0} = a^{K/4\pi+\Delta_{\hat{\phi}}-2}g\,.
\end{equation}
We proceed to the higher-order calculation of the RG. In the following calculation, we perturb near $K_c = 2-\Delta_{\hat{\phi}}$ and define $\delta = K/4\pi-2+\Delta_{\hat{\phi}}$. The log divergence in the 1PI diagrams for the $\varphi$ propagator in Fig.~\ref{fig:hofd} (b).  
The self-energy of $\sqrt{K}\varphi$ takes the form
\begin{equation}
    \Sigma_\varphi^{(1)}(p)  = 0 \,,
\end{equation}
\begin{equation}
\begin{aligned}
    \Sigma^{(2)}_\varphi(p)&  = -\frac{g_\text{bare}^2}{K} \int d^2x \left(e^{i p x} G_\phi(x) \sinh G_\varphi(x)-G_\phi(x) \cosh G_\varphi(x)\right)\\
    & \sim -\frac{1}{2}(1+2\delta\log a) \int d^2x \left(e^{i p x}-1\right) e^{G_{\varphi}(x)} G_\phi(x)\\
    & = -\frac{1}{4}\frac{1}{4\pi(2-\Delta_{\hat{\phi}})}(1-\frac{1}{2-\Delta_{\hat{\phi}}}\delta+2\delta\log a)g^2 \int dx\frac{2\pi p^2x^3}{(x^2+a^2)^{2+\delta}} \\
    & \sim \frac{g^2p^2}{8(2-\Delta_{\hat{\phi}})}\left[1-\left(1+\frac{1}{2-\Delta_{\hat{\phi}}}\right)\delta\right]\log a\,.
    \end{aligned}
\end{equation}
By requiring 
\begin{equation}
    G_\varphi(p)^{-1} = Z_\varphi p^2+\Sigma_{\varphi}(p)
\end{equation}
is finite, we extract $Z_{\varphi}$ factor
 \begin{equation}
     Z_{\varphi} = 1-\frac{g^2}{8(2-\Delta_{\hat{\phi}})}\log a+\frac{(3-\Delta_{\hat{\phi}})g^2\delta}{8(2-\Delta_{\hat{\phi}})^2}\log a,
 \end{equation}
Using Eq.~\eqref{Zfactor}, we have the beta function of $\delta$
\begin{equation}
    \beta_\delta = (1+\delta)\frac{\partial Z_{\varphi}}{\partial \log a} = -\frac{g^2}{8(2-\Delta_{\hat{\phi}})}+\frac{g^2\delta}{8(2-\Delta_{\hat{\phi}})}.
\end{equation}
We now implement the calculation of beta function of $g$ up to $\mathcal{O}(g^3)$. 
The higher-order diagrams that have the proper log divergence are illustrated in Fig.~\ref{fig:hofd} (c). 
They take the form of
 \begin{equation}
     \Gamma_{g}^{(1)}(0) = g_{\text{bare}}\,,
 \end{equation}
 \begin{equation}
     \Gamma_{g}^{(2)}(0) = 0\,,
 \end{equation}
 \begin{equation}\label{eq::gamma3}
     \begin{aligned}
         &\Gamma_{g}^{(3)}(0)  \sim \frac{g_{\text{bare}}^3}{K^2}\int d^2 x d^2 y[ \sinh G_\varphi(x-y)(\cosh G_\varphi(y)-1) \cosh G_\varphi(x) G_\phi(x)+\sinh G_\varphi(x-y) \cosh G_\varphi(y) G_\phi(y)(\cosh G_\varphi(x)-1) \\
        &\left.+\sinh G_\varphi(x-y) G_\phi(x-y)(\cosh G_\varphi(x)-1)(\cosh G_\varphi(y)-1)\right]\\
        &-\frac{g_{\text{bare}}^3}{K^2} \int d^2 x d^2 y\left[(\cosh G_\varphi(x-y)-1) \sinh G_\varphi(x) G_\phi(x) \sinh G_\varphi(y)+(\cosh G_\varphi(x-y)-1) \sinh G_\varphi(x) \sinh G_\varphi(y) G_\phi(y)\right. \\
        &\left.+\cosh G_\varphi(x-y) G_\phi(x-y) \sinh G_\varphi(x) \sinh G_\varphi(y)\right]\\
        &-\frac{1}{2} \frac{g_{\text{bare}}^3}{K^2} \int d^2 x d^2 y[(\cosh G_\varphi(x-y)-1)(\cosh G_\varphi(y)-1) \cosh G_\varphi(x) G_\phi(x)\\
        &+(\cosh G_\varphi(x-y)-1)(\cosh G_\varphi(x)-1) \cosh G_\varphi(y) G_\phi(y) \left.+(\cosh G_\varphi(x)-1)(\cosh G_\varphi(y)-1) \cosh           G_\varphi(x-y) G_\phi(x-y)\right]\\
        & \frac{1}{2} \frac{g_{\text{bare}}^3}{K^2} \int d^2 x d^2 y\left[\sinh I(x-y) \sinh I(x) G_\phi(x) \sinh I(y)+\sinh I(x-y) \sinh I(x) \sinh I(y) G_\phi(y)\right. \\
        &\left.+\sinh I(x-y) G_\phi(x-y) \sinh I(x) \sinh I(y)\right]\,
\end{aligned}
\end{equation}
The summation of the terms that have the proper log divergence is
\begin{equation}
    \frac{1}{4}\frac{(g_\text{bare})^3}{K^2}  \int d^2xd^2y G_\phi(x) e^{G_\varphi(x)+G_\varphi(y)-G_{\varphi}(x+y)}
\end{equation}
where we have used the exchange symmetry $x\leftrightarrow y$ and $y\leftrightarrow -y$ in the integral. 
The integral is divided into three parts:
\begin{equation}
    \text{Region 1}: |y|,|x|<\Delta \,, \quad \text{Region 2}: |x|<\Delta,|y|>\Delta\,, \quad \text{Region 3}: |x|>\Delta,|y|<\Delta\,.
\end{equation}
We first perform the evaluation of the integral in the first region. In region I, we can approximate both $G_\varphi(x)$, $G_{\varphi}(y)$ and $G_{\varphi}(x+y)$ into the log function, and directly work out the integral
\begin{equation}
\begin{aligned}
    J_1 & = \frac{1}{4}\frac{(g_\text{bare})^3}{K^2}  \int_{\mathcal{R}_1} d^2xd^2y G_\phi(x) e^{G_\varphi(x)+G_\varphi(y)-G_{\varphi}(x+y)} \\
    & = \frac{1}{4}\frac{(g_\text{bare})^3}{K^2}\int_{\mathcal{R}_1} d^2 x d^2 y \frac{\left((x-y)^2+a^2\right)^{2-\Delta_\phi}}{\left(x^2+a^2\right)^2\left(y^2+a^2\right)^{2-\Delta_\phi}}\\
    & \sim \frac{1}{64} g^3\left(\log \left(\frac{a}{\Delta}\right)^2+\log ^2\left(\frac{a}{\Delta}\right)^2\right)\,.
    \end{aligned}
\end{equation}
The integral at the second line cannot explicitly be worked out for general $\Delta_{\hat{\phi}}$, hence we would approximate $\Delta_{\hat{\phi}}  \approx 1$ because the estimate of $\Delta_\text{ord} $ from Monte Carlo simulation is $\Delta_\text{ord} \approx 1.2626$.
The contribution from region 2 is identical to region 3, and we perform the Taylor expansion for small $x$, say $G_{\varphi}(x+y)=G_{\varphi}(y)+h G_{\varphi}^{\prime}(y)+\frac{1}{2} h^2 G_\varphi^{\prime \prime}(y)$, where $h = 2xy\cos\theta+x^2$. 
The integrals are evaluated to be 
\begin{equation}
\begin{aligned}
    J_2 = J_3 & = \frac{1}{4} \frac{\left(g_{\mathrm{bare}}\right)^3}{K^2} \int_{\mathcal{R}_2} d^2 x d^2 y G_\phi(x) e^{G_{\varphi}(x)+G_{\varphi}(y)-G_{\varphi}(x+y)}\\
    & = \frac{\pi}{2}\frac{g_\text{bare}^3}{K^2} \int_0^{\Delta} d^2 x x^2 G_\phi(x) e^{G_\varphi(x)} \int_{\Delta}^{\infty} yd y \int_0^{2 \pi} d \theta \left(-G_\varphi^{\prime}(y)+2 y^2 \cos \theta^2 G_\varphi^{\prime}(y)^2-2 y^2 \cos \theta^2 G_\varphi^{\prime \prime}(y)\right)\\
    & \sim -\frac{1}{64}g^3 \log \left(\frac{a^2}{\Delta^2}\right) \int_{\Delta}^{\infty} d y\left(2 y K_0^{\prime}(y)^2-y K_0^{\prime \prime}(y)\right)\\
    &  = -\left.\frac{1}{64} g^3 \log \left(\frac{a^2}{\Delta^2}\right)\left(y^2\left(K_1^2-K_0 K_2(y)\right)+ y K_1(y)\right)\right|_{\Delta} ^{\infty}\\
    & \sim \frac{1}{64} g^3 \log \left(\frac{a^2}{\Delta^2}\right)\left(2+2\log \left(\Delta\right)\right)\,.
    \end{aligned}
\end{equation}
Summing over contributions from all regions, the divergence is
\begin{equation}
    J = J_1+J_2+J_3 \sim \frac{5}{32}g^3\log a+\frac{1}{16}g^3\log^2 a\, .
\end{equation}

\begin{figure}[t]
    \centering
\includegraphics[width = 0.42\textwidth]{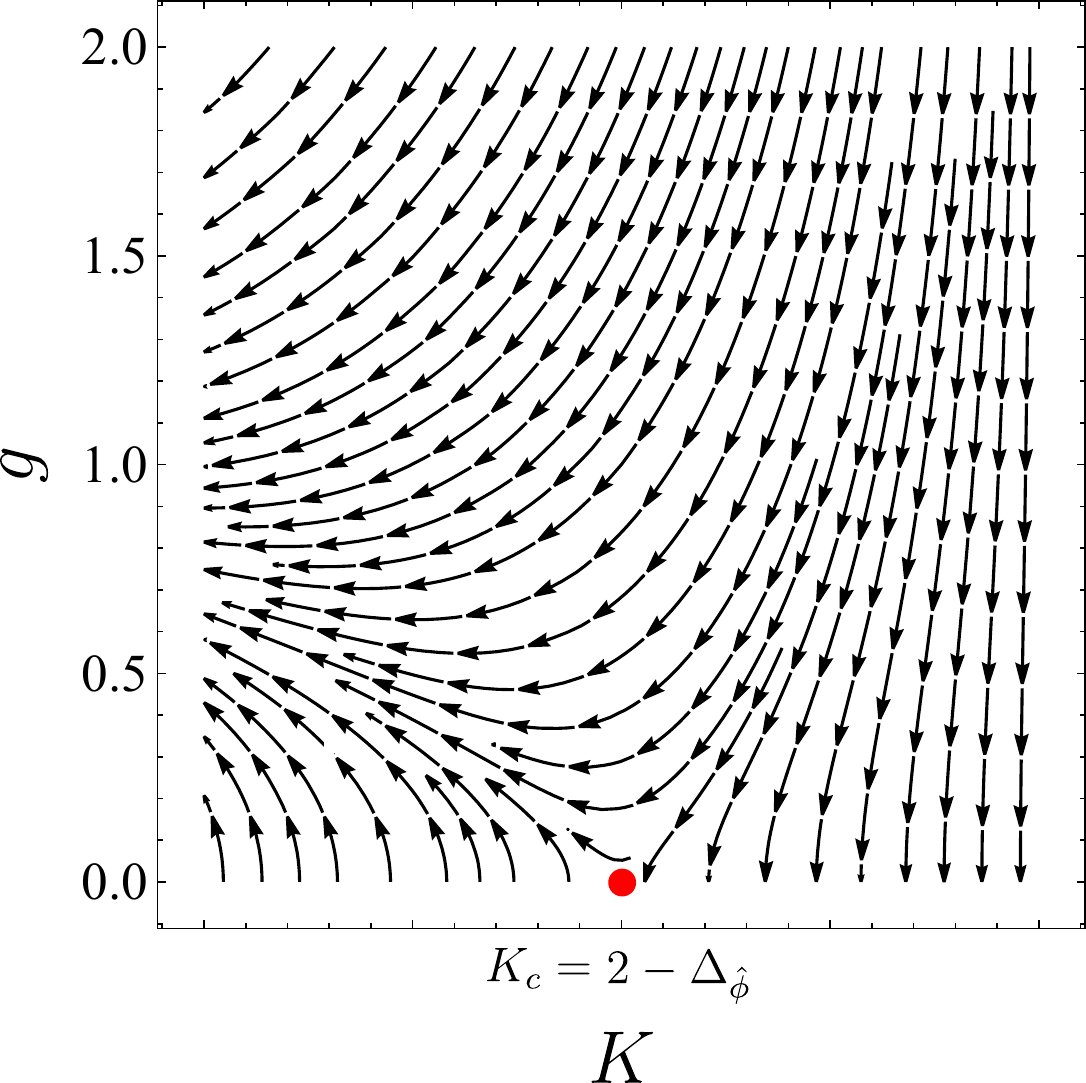}
    \caption{Flow pattern of the special BKT phase transition in from the higher order RG equation.}
    \label{fig:hosbkt}
\end{figure}

Hence the $Z_g$ is 
\begin{equation}
    Z_g = 1+\frac{5}{32}g^3\log a\, .
\end{equation}
Finally, the higher order beta function for $\delta$ and $g$ is
\begin{equation}
    \begin{aligned}
        \beta_\delta & = -\frac{g^2}{8}+\frac{g^2\delta}{8} \,,\\
        \beta_g & = -\delta g-\frac{5}{32(2-\Delta_{\hat{\phi}})^2}g^3 \,.
    \end{aligned}
\end{equation}
Recall that $\delta = K/4\pi-2+\Delta_{\hat{\phi}}$, by rescaling $K/4\pi\to K, g^2/8\to g^2$, the flow pattern at the $K-g$ plane is shown in Fig.~\ref{fig:hosbkt}, at the strongly coupled region, the RG flows to $K = 0$, which corresponds to the gapped phase.

\end{document}